\documentclass[acmtog]{acmart}

\acmJournal{TOG}
\copyrightyear{2026}
\acmYear{2026}
\setcopyright{cc}
\setcctype{by}
\acmConference[SA Conference Papers '26]{SIGGRAPH Asia 2026 Conference Papers}{December 01--04, 2026}{Kuala Lumpur, Malaysia}
\acmBooktitle{SIGGRAPH Asia 2026 Conference Papers (SA Conference Papers '26), December 01--04, 2026, Kuala Lumpur, Malaysia}
\acmDOI{10.1145/3829340.3842259}
\acmISBN{979-8-4007-2842-6/2026/12}

\AtBeginDocument{%
  }

\usepackage{bm}
\usepackage{makecell}
\usepackage{multirow}
\usepackage{subcaption}

\usepackage{acmart-taps}

\newcommand{\Cue}{\textit{Cue Sequence }}

\newcommand{\Dataset}{\textbf{Musilux}}
\newcommand{\Corpus}{\textit{Cue Sequence Corpus }}
\newcommand{\LAMP}{\textbf{LAMP }}

\begin{document}

%% Title (with short title for running head).
\title[AuraLuxMuse]{AuraLuxMuse: Adaptive Fusion Modeling for
  Aesthetic Stage Lighting Design with Music and Expert Guidance}

%% Authors and affiliations (ACM format).
\author{Junyu Deng}
\orcid{0009-0004-6313-7378}
\affiliation{%
  \institution{Fudan University}
  \city{Shanghai}
  \country{China}}
\email{21307140005@m.fudan.edu.cn}

\author{Jiale Cao}
\orcid{0009-0001-8635-4306}
\affiliation{%
  \institution{Fudan University}
  \city{Shanghai}
  \country{China}}
\email{23300240001@m.fudan.edu.cn}

\author{Mengtian Li}
\orcid{0009-0001-7281-2812}
\authornote{Mengtian Li and Zuo Hu are the corresponding authors.}
\affiliation{%
  \institution{Fudan University}
  \city{Shanghai}
  \country{China}}
\affiliation{%
  \department{Shanghai Film Academy}
  \institution{Shanghai University}
  \city{Shanghai}
  \country{China}}
\email{mtli@fudan.edu.cn}

\author{Zhongxia Ji}
\orcid{0009-0005-9550-8548}
\affiliation{%
  \institution{Shanghai Theatre Academy}
  \city{Shanghai}
  \country{China}}
\email{zhongxia.ji@sta.edu.cn}

\author{Ruhua Chen}
\orcid{0009-0003-0619-765X}
\affiliation{%
  \institution{Shanghai Theatre Academy}
  \city{Shanghai}
  \country{China}}
\email{janechen2-c@my.cityu.edu.hk}

\author{Yiyi He}
\orcid{0009-0003-3348-6764}
\affiliation{%
  \institution{Nanjing University of the Arts}
  \city{Nanjing}
  \country{China}}
\email{125751790@qq.com}

\author{Guangnan Ye}
\orcid{0009-0007-4973-7942}
\affiliation{%
  \institution{Fudan University}
  \city{Shanghai}
  \country{China}}
\email{yegn@fudan.edu.cn}

\author{Zuo Hu}
\orcid{0009-0009-9272-8560}
\authornotemark[1]
\affiliation{%
  \institution{Shanghai Theatre Academy}
  \city{Shanghai}
  \country{China}}
\email{huzuo@sta.edu.cn}

%% Abstract.
\begin{abstract}
We present \textbf{AuraLuxMuse}, a novel system for automated aesthetic stage lighting design that integrates expert knowledge, representation learning, and preference-adaptive modeling. Lighting design in live performance settings requires the seamless translation of musical features into dynamic lighting behaviors. However, traditional workflows remain time-consuming, labor-intensive, and difficult to transfer.
AuraLuxMuse encodes music and professional cue sequences into a shared retrieval space, estimates cue-event density, and retargets selected fixture commands to the destination stage. It assists pre-production authoring by returning editable cues rather than replacing the designer with an unconstrained generator.
At the heart of AuraLuxMuse are two key modules: Lighting-Aligned Music Pretraining (\textbf{LAMP}), which performs contrastive learning between audio and lighting cues for alignment, and Preference-Adaptive Mixture of Experts (\textbf{PAMoE}), which conditions preference-aware cue retrieval and adaptation on designers' intent through a gated ensemble of style-specific expert networks. To support training and evaluation, we introduce \Dataset, the first dataset of paired musical audio and professional lighting cue sequences under diverse performance scenarios. We evaluate AuraLuxMuse across both virtual simulation environments and professional-grade laboratories. Experimental results, including objective and subjective evaluation, demonstrate that AuraLuxMuse retrieves and adapts stage-lighting cues that are visually cohesive, semantically meaningful, and artistically expressive, showing its potential for AI-assisted aesthetic stage design.
\end{abstract}

%% CCS concepts (from old.tex; generated via dl.acm.org/ccs).


%% Keywords (from old.tex).

%% Teaser figure (uses the teaserfigure environment provided by acmart).
\begin{teaserfigure}
  \includegraphics[width=\textwidth]{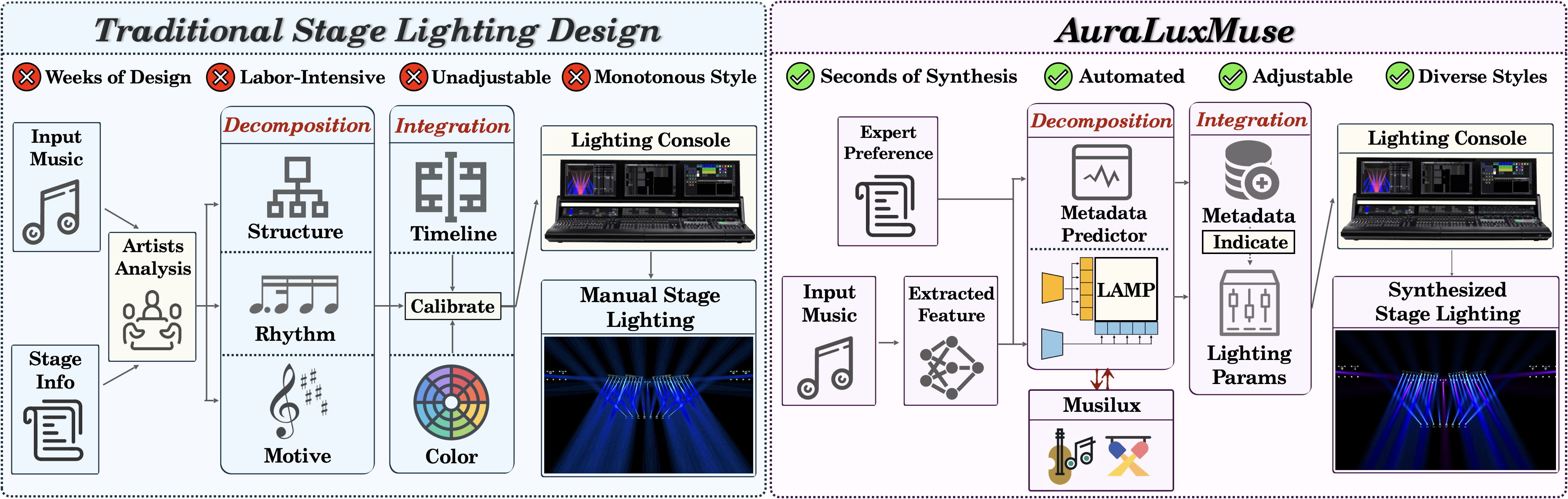}
  \caption{\textbf{Procedure Comparison: Traditional Stage Lighting Design vs.\ AuraLuxMuse.}~Conventional pipelines require designers to manually decompose music into artistic components, then iteratively design cues, align timing, and input parameters into consoles. AuraLuxMuse accelerates this workflow through feature extraction, preference-conditioned retrieval, cue adaptation, and console dispatch.}
  \Description{Comparison of a traditional manual stage-lighting workflow with the AuraLuxMuse workflow, from music input and artistic analysis to console-ready lighting output.}
  \label{fig:teaser}
\end{teaserfigure}

\maketitle

\section{Introduction}
Stage lighting, as defined by professional lighting designer R. Craig Wolf~\cite{scene_design}, refers to the art of using lighting techniques to illuminate the stage and shape the appearance of characters and scenery. It fulfills the artistic vision of designers by translating abstract concepts such as atmosphere, rhythm, and emotion into dynamic visual compositions in live performance. Stage lighting plays a vital role in the overall stage experience, engaging audiences through carefully orchestrated lighting behaviors.
At the heart of it lies stage lighting design, which encompasses fixture behavior programming and real-time adjustment. An aesthetic lighting design must be technically feasible, artistically expressive, and adaptable to various performance scenarios.

Traditional lighting design~\cite{scene_light, light_space} begins with meticulous analysis of musical structure: segmenting phrases, detecting transitions, and extracting rhythmic elements. Crucially, designers must interpret the musical motive into a thematic concept that drives the formation of lighting behaviors, guiding the programming of individual fixtures that are aggregated into control parameters. This process is highly manual and iterative, often requiring multiple revisions informed by visual feedback, as illustrated in Fig.~\ref{fig:teaser}. While this handcrafted alignment of music, space, and light can yield emotionally resonant effects, it is inherently time-consuming and labor-intensive. Each lighting cue must be programmed from scratch, leading to low reusability and limited scalability. Moreover, aesthetic quality and structural consistency vary significantly between productions.

Consequently, there is a compelling need for a system capable of modeling and automating the entire stage lighting design process. However, existing machine-learning systems typically restrict lighting control to basic properties such as light position and intensity, overlooking critical expressive dimensions including color, stylistic variation, and temporal rhythm. To the best of our knowledge, this topic has not been systematically explored under modern deep-learning paradigms. With advances in deep audio representation learning, it is now feasible to encode the latent structure of music in high-dimensional embedding spaces, providing a promising foundation for music-driven lighting synthesis. Three key challenges remain. The first is \textbf{lack of structured parameterization}: there is no standardized formulation for decomposing lighting attributes into learnable parameters. The second is \textbf{multimodal alignment and personalization}: learning effective correspondences between music and lighting remains difficult, particularly when integrating personalized and stylistic preferences. The third is the \textbf{absence of comprehensive evaluation metrics}: existing benchmarks fail to assess lighting quality in terms of both objective fidelity and subjective perception.

To address these challenges, we propose \textbf{AuraLuxMuse}, a novel system integrating expert knowledge, multimodal representation, and preference-adaptive modeling for stage lighting design.
As illustrated in Fig.~\ref{fig:teaser}, the pipeline operates directly on music and captures intrinsic properties. It uses these properties and a designer-specified preference to retrieve professional cue sequences, retargets their fixture commands to the destination stage, and dispatches the resulting cues and timecodes to the lighting console.

Specifically, AuraLuxMuse comprises two core modules: Lighting-Aligned Music Pretraining (\textbf{LAMP}) and Preference-Adaptive Mixture of Experts (\textbf{PAMoE}).
At the core of cross-modal alignment, LAMP bridges the semantic gap between music and lighting. It encodes both audio and lighting parameters into a shared embedding space. This enables the system to retrieve the most semantically aligned lighting cues from a corpus, tailored to the corresponding musical content.
To support stylistic control, PAMoE combines four learned MLP branches using both cue-dependent and preference-dependent gate weights. The branches are not preassigned to individual style axes; their observed associations with Color, Brightness, Zone, and Rhythm are analyzed after training. The resulting representation steers retrieval toward cues consistent with the designer's selected preference.

A professional lighting program comprises fixture-level \textit{Cue} commands covering attributes such as intensity, color, pan, tilt, and gobo values, together with \textit{Timecodes} that trigger cue sequences on a console. Given an input music track and preference vector $s=[s_c,s_b,s_z,s_r]$, the system encodes both as a retrieval query. The Metadata Predictor estimates the required cue density $K$, enabling the retriever to select the top-$K$ professional cue sequences. Stage patching converts them into console-ready files.

We further introduce \Dataset, a temporally aligned dataset containing 187 minutes and 22 seconds of music and more than 80,000 professional lighting instructions across real-world performance scenarios.
Our key contributions are:
\begin{itemize}
  \item We introduce \textbf{AuraLuxMuse}, a preference-conditioned retrieval and adaptation system whose outputs are competitive with manual designs in the evaluated scenarios, and \Dataset, the first dataset to explicitly link professional lighting control parameters with musical audio.
  \item We develop \LAMP for cross-modal learning between lighting and music, \textbf{PAMoE} for stylistic revision based on designers' intent, and evaluation diagnostic \textbf{EGSS} for retrieved and adapted stage-lighting outputs.
  \item We deploy AuraLuxMuse on virtual and real-world stages, producing editable candidate cues with an average inference time of \textbf{$7.65 \pm 0.98$ seconds} under the reported setup, as supported by qualitative and quantitative evaluations.
\end{itemize}

\section{Related Work}
\paragraph{Music-Driven Motion, Visual Generation, and Lighting Design}

Music-conditioned synthesis has been explored for dance and articulated motion, including long-horizon 3D choreography from paired music-motion data~\cite{aistpp_choreographer}. Related audio-to-visual work predicts visual dynamics from sound and context~\cite{sound2sight} or generates human-centric video conditioned on music~\cite{music_conditioned_video}. These tasks synthesize pixels or body motion and do not produce fixture-addressed, console-executable lighting cues.
Lighting research instead spans rendering, design assistance, and illumination control, including simulation-based methods, natural illumination~\cite{envylight}, stage-lighting systems~\cite{interactive, simulation, Exploratory}, virtual entertainment~\cite{toward, paint, automating}, architectural and indoor illumination~\cite{lightguider, data-driven, indoor, layout}, and image-based relighting~\cite{dilight, photography}. Existing lighting systems mainly formulate the problem as interactive design, physically based simulation, or objective-driven optimization, typically requiring designer-specified objectives, manual control, or predefined programs.
To the best of our knowledge, prior work does not retrieve fixture-addressed, console-executable professional cue sequences directly from music while also conditioning the retrieval on designer preference and retargeting it to a new stage.
Retrieval-Augmented Generation (RAG)~\cite{rag} is a powerful paradigm for injecting external knowledge or examples into generative models, with extensions beyond text: WavRAG~\cite{wavrag} targets spoken dialogue, and TTA-RAG~\cite{ttarag} improves text-to-audio synthesis in few-shot regimes.

Retrieval-based creative authoring reuses curated assets or examples rather than synthesizing every low-level element from scratch. RAG and its audio extensions~\cite{rag,wavrag,ttarag} provide the general retrieval paradigm; AuraLuxMuse applies the idea to a structured repository of expert-authored, console-valid cue sequences. We therefore cast stage-lighting design as preference-conditioned cross-modal retrieval and adaptation: musical context and user-specified expressive intent jointly query the cue repository for cues that are temporally coherent, rhythmically grounded, and stylistically aligned.
% These efforts demonstrate previous exploration on lighting design task and the effective retrieval-enhanced modeling capabilities in multimodal understanding and generating systems. The previous works on stage lighting often lack scalability to real-world scenarios and generalization capacity, or are only applicable to static scenarios rather than dynamic ones. As a result, we established a deep learning based pipeline to tackle the shortcomings.
\paragraph{Audio Alignment}
Convolutional Neural Networks have significantly advanced audio processing \cite{cnn}. MFCC remains a foundational time-frequency representation \cite{MFCC}.
Among CNN-based architectures, VGGish \cite{vggish} is widely adopted for its transferable features, while transformer-based models such as SpecTNT \cite{transformermusic} and AST \cite{ast} have demonstrated strong capacity for modeling spectrotemporal dependencies.
Beyond unimodal learning, deep spectrum translation \cite{kashani2019speech} and audiovisual localization \cite{chen2024unveiling} have broadened multimodal artistic applications. Other methods enhance robustness across domains \cite{wu2025complexdec}.
Contrastive learning has emerged as a powerful approach for cross-modal representation, with CLIP~\cite{clip} marking a key milestone.
CLaMP~\cite{clamp} aligns symbolic music with language to enable robust music retrieval, motivating extensions such as AudioCLIP~\cite{audioclip}, CLAP~\cite{CLAP}, SLAP~\cite{SLAP} and MusCALL~\cite{manco2022} to extend this paradigm to audio, supporting the retrieval of multiple modalities. CLMR~\cite{spijkervet2021} leverages self-supervised contrastive learning on unlabeled data.

Modality alignment is essential for cross-domain design. 
We use audio as the guiding modality and extract high-level musical representations to model cross-modal correspondence. 
Inspired by contrastive multimodal learning, we introduce an alignment module that maps music and lighting into a shared latent space, enabling coherent interactions across modalities.
\paragraph{Designer Preference Adaptation} The Mixture of Experts (MoE) architecture~\cite{jacobs1991adaptive} has gained widespread adoption because it decomposes complex demands into specialized subtasks. Sparse MoE layers have demonstrated scalability in NLP~\cite{fedus2022switch,shazeer2017outrageous} and image synthesis~\cite{lee2018megan}. Variational MoE autoencoders~\cite{autoencoders} have also been introduced for cross-modal generation. Furthermore, MoE has been adopted for audiovisual learning~\cite{cheng2024mixtures} and style-specific text-to-speech synthesis~\cite{stylemixtureexperts}, underscoring its potential in tasks requiring expressive generation.

As a creative task, stage lighting design inherently reflects the stylistic preferences of artists, necessitating a mechanism for incorporating personal preference and adapting model behavior accordingly. To this end, MoE provides a natural solution, enabling the model to dynamically select specialized expert pathways.

\section{Method}
\subsection{System Overview and Problem Formulation}
AuraLuxMuse is designed for offline pre-production, where artists can interact with and fine-tune stage-lighting effects.
Its input is a music track together with an optional four-axis preference vector.
Its output is a set of editable \textit{Cue Sequences} and \textit{Timecodes}, not rendered pixels.
For example, given a concert excerpt and a high-rhythm preference vector, LAMP retrieves temporally compatible professional cues, the Metadata Predictor determines how many spatially complementary sequences are needed, PAMoE changes the preference-conditioned cue representation, and the patching procedure converts fixture identifiers and attributes to the target console layout.
\label{sec:problem_statement}
The objective is to retrieve and adapt temporally aligned professional lighting events conditioned on music and stylistic preference.
Given a musical performance represented as a continuous feature sequence $M=\{M_t\}_{t=1}^{T}$ and an optional stylistic preference vector $s \in \mathbb{R}^{d_s}$, our objective is to generate a lighting sequence $L = \{ (T_i, CS_i) \}_{i=1}^{N}$, where $T_i \in \mathbb{R}^{+}$ denotes the timecode at which a lighting cue is triggered, $CS_i \in \mathbb{R}^{d_c}$ represents the corresponding cue parameters retrieved from a collected cue corpus $C$, and $N$ denotes the estimated total number of lighting events.
% Furthermore, according to leading scholars in stage lighting design, automating such an inherently artistic process introduces several technical and aesthetic challenges, including the decomposition of lighting parameters, cross-modal alignment between auditory and visual domains, coordination of temporal event density, and preservation of stylistic distinctiveness.

\begin{figure*}[ht]
  \centering
  \includegraphics[width=\linewidth]{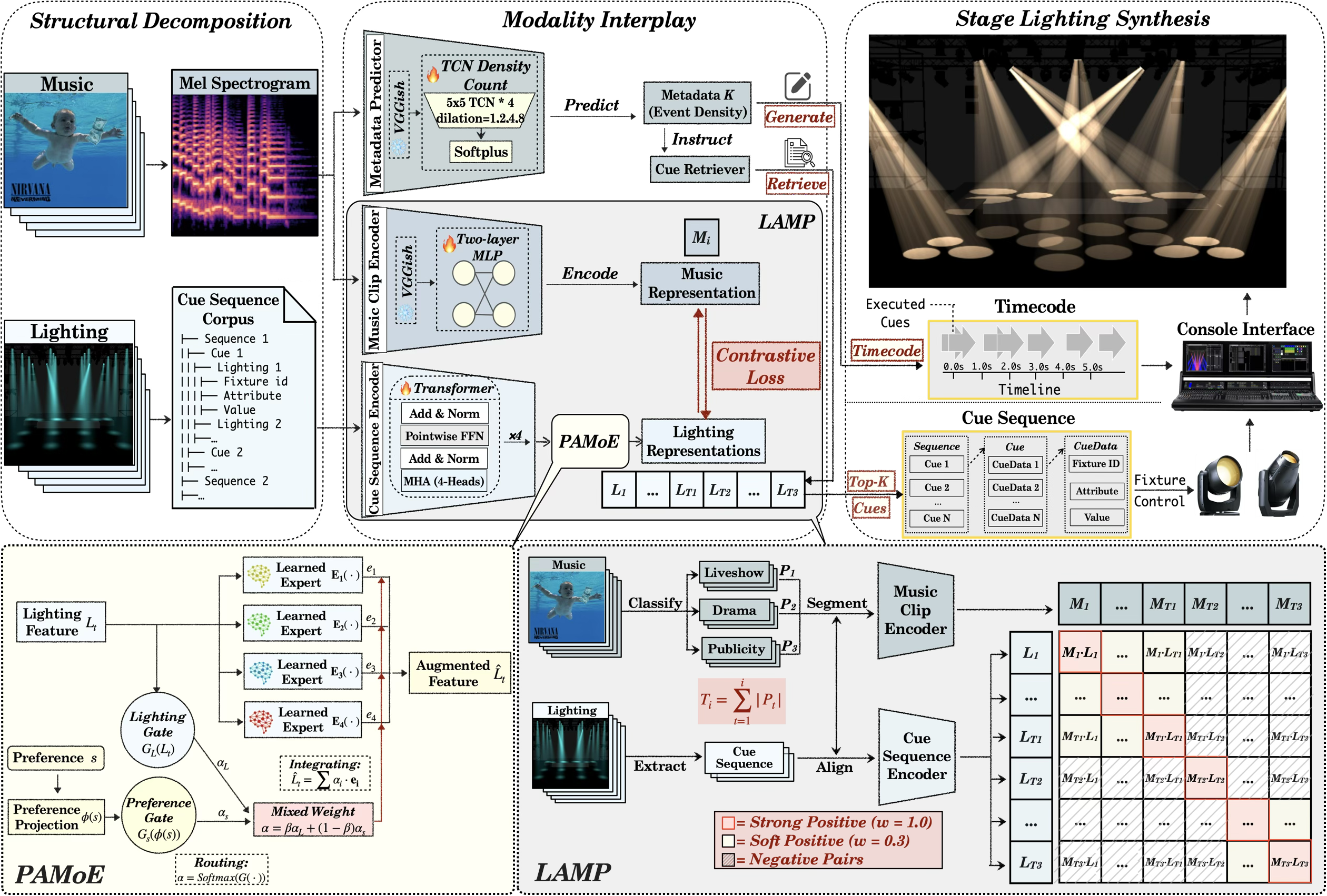}
  \caption{\textbf{Overview of the AuraLuxMuse Pipeline.}
  Music and lighting data are individually preprocessed before entering LAMP for contrastive learning, and preference representations are injected via PAMoE. $P_1/P_2/P_3$ denote per-style frame counts, and $T_i=\sum_{t=1}^{i}|P_t|$ is the cumulative sample count used to index the illustrated segments. Cosine similarity and the metadata value $K$ predicted by the non-causal TCN Metadata Predictor are used to query the cue corpus.}
  \Description{AuraLuxMuse pipeline showing music and cue encoders, LAMP contrastive alignment, PAMoE preference routing, the non-causal TCN Metadata Predictor, top-K cue retrieval, and console execution.}
  \label{fig:pipeline}
\end{figure*}

\subsection{Structural Decomposition: Cues \& Timecodes}
\label{sec:decomposition}
% \textit{Stage Lighting} represents the art of illuminating theater, opera, and other performance arts to enhance visibility, mood, and narrative impact. 
The principled computational representation of stage lighting remains unexplored, with little prior work on decomposing its control parameters. This remains as a challenge.
To tackle this, we systematically investigate the decomposition of stage lighting and propose a structured breakdown that formalizes computational representation.
In industrial systems, stage lighting is parameterized through two fundamental components: \Cue and \textit{Timecodes}. Together, they orchestrate precisely synchronized lighting across performance. A \textit{Cue Sequence} is defined as a structured list of lighting directives, containing cues that govern the behavior of each lighting fixture.
\textit{Timecodes} serve as the temporal control system for cue execution. Each cue is annotated with a corresponding timestamp, ensuring the lighting transitions are executed with frame-level precision.
A Cue Sequence comprises multiple Cues, each containing CueData units that encode control parameters for individual lighting fixtures.
Each Timecode entry specifies the exact moment at which a particular Cue Sequence should be triggered.
This data structure serves as the foundation underlying our proposed dataset detailed in Section~\ref{sec:dataset}.
% This data structure, as depicted in Fig.~\ref{fig:data_structure}, serves as the foundation underlying our proposed dataset detailed in Section~\ref{sec:dataset}.

% \subsection{Contrastive Learning for Modal Fusion: LAMP}
\subsection{Modality Interplay via Alignment: LAMP}
The interplay between sonic and visual elements generates a synaesthetic experience in which light can be perceived as having acoustic properties and vice versa.

We formulate this relationship as a cross-modal contrastive learning problem. The interplay between music and lighting is fundamental to artistic expression. We therefore introduce Lighting-Aligned Music Pretraining (\textbf{LAMP}), an architecture that represents lighting features and their corresponding music clips in a shared embedding space, as depicted in Fig.~\ref{fig:pipeline}.
% , together with the similarity matrix in Fig.~\ref{fig:LAMP_table}.
% To avoid challenges such as tedious strategy design under embedding procedure and limited interpretability of generative approaches, the idea of contrastive learning for embedding is then applied, offering a more scalable and explainable solution for multimodal alignment.
We use only frames where lighting fixtures are active and construct each LAMP batch from paired raw music windows $\widetilde{M}_i$ and temporally matched raw cue-sequence segments $\widetilde{L}_j$, yielding $M_i=f_M(\widetilde{M}_i)$ and $L_j=f_L(\widetilde{L}_j)$ from the Music Clip Encoder and Cue Sequence Encoder.
Beyond exact same-music/same-moment pairs, we include positive pairs defined by the scenario and style labels described in Section~\ref{sec:data_collection}.
By applying weighted alignment across these multiple positive matches, we optimize the contrastive loss in Equation~(\ref{contrastive_loss}).
\begin{equation}
\mathcal{L} = -\frac{1}{N} \sum_{i=1}^{N} \log \frac{\sum\limits_{p \in \mathcal{P}(i)} w_{i,p} \cdot \exp\left( \frac{\operatorname{sim}(M_i, L_p)}{\tau} \right)}{\sum\limits_{j=1}^{N} \exp\left( \frac{\operatorname{sim}(M_i, L_j)}{\tau} \right)}.
\label{contrastive_loss}
\end{equation}

\noindent Here, $\operatorname{sim}(M_i,L_j)$ is cosine similarity in the shared embedding space.
\begin{equation}
\operatorname{sim}(M_i,L_j)=\frac{M_i^\top L_j}{\lVert M_i\rVert_2\,\lVert L_j\rVert_2}.
\end{equation}
The functions $f_M$ and $f_L$ are the music and cue-sequence encoders. The positive set $\mathcal{P}(i)$ contains the original music--lighting pair as a strong positive ($w_{i,p}=1$) and same-style, same-music segments from different moments as soft positives ($w_{i,p}=0.3$). Here, $N$ is the batch size and $\tau=0.07$ is the softmax temperature; Supplementary Section~E examines the weights.
% The model is optimized via the proposed loss function and maximizes similarity for positively aligned pairs while minimizing it for negatives. %Consequently, LAMP learns a semantically meaningful embedding space for downstream operations.
\subsection{Frame-wise Estimation: Metadata Predictor}

The density of lighting events is directly related to the rhythmic pace of performances. Designers must calculate the appropriate frequency of lighting changes to support the narrative without overwhelming the audience.

To address, we propose a Metadata Predictor that estimates the expected number of lighting events in each temporal window, denoted by the metadata value $K$. This module learns a non-negative frame-level density, inspired by the crowd-counting paradigm~\cite{cross_scene, count_object}, in which lighting events are represented by density values rather than discrete timestamps.
Formally, let $M_{s:s+W-1}$ denote a music-feature window of $W$ frames beginning at frame $s$. The predictor outputs a density vector $\boldsymbol{\lambda}_s$, whose sum gives the estimated event count for that window:
\begin{align}
\boldsymbol{\lambda}_s &= \mathrm{Softplus}\!\left(\mathrm{TCN}_{\phi}\!\left(M_{s:s+W-1}\right)\right),\\
\widehat{c}_s &= \sum_{\tau=0}^{W-1}\lambda_{s,\tau},
\qquad K_s = \left\lfloor \widehat{c}_s \right\rceil.
\end{align}

The estimator runs offline and uses a non-causal TCN with symmetric padding. Consequently, an interior frame can use both preceding and subsequent context within the complete input window before the predicted density values are summed. The TCN uses dilated one-dimensional kernels, and Softplus enforces non-negativity. The window length $W$, stride, and learning rate are summarized in Section~\ref{sec:setup}.
This formulation offers three notable advantages aligned with stage-lighting prediction. It mitigates data sparsity by converting extremely sparse frame-wise labels into smooth temporal supervision, improving stability and optimization. It maintains physical validity because enforcing $\lambda_{s,\tau}\ge0$ ensures non-negative event predictions. It also provides temporal awareness because window-level supervision over $W$ frames captures short-term music--lighting dependencies.

\subsection{Mechanism for Artists' Specific Needs: PAMoE}
Each lighting designer develops a recognizable visual vocabulary, so stage lighting embodies personal style and preference that must be explicitly modeled in any computational formulation.

To address this requirement, we incorporate personalized stylistic control into the system.
In collaboration with professional lighting designers, we identify and annotate four ordinal stylistic dimensions: Color $c$, Brightness $b$, Zone $z$, and Rhythm $r$, with label definitions detailed in Supplementary Section~C. The preference vector $s=[s_c,s_b,s_z,s_r]$ is used during joint LAMP--PAMoE training and can also be adjusted by a designer at inference to steer cue retrieval. To integrate this control signal, we propose Preference-Adaptive Mixture of Experts (\textbf{PAMoE}).
As illustrated in Fig.~\ref{fig:pipeline}, PAMoE is a dense soft mixture of four architecturally symmetric MLP branches, denoted as Learned Experts $E_1$--$E_4$. All four branches are evaluated for each lighting sequence. The branches are not hard-assigned to exact style dimensions. Specialization is learned jointly and reported through the gate-sensitivity correlations in Supplementary Section~C.
Given a lighting-feature sequence $L=\{L_t\}_{t=1}^{n}\in\mathbb{R}^{n\times d}$ and preference vector $s=[s_c,s_b,s_z,s_r]^\top$, each ordinal value is represented by a 32-D embedding and the four vectors are concatenated into the 128-D preference projection $\phi(s)$. The Lighting Gate first performs sequence mean pooling and produces the data-dependent distribution $\alpha_L=\operatorname{softmax}(G_L(L))$. Independently, the Preference Gate produces $\alpha_s=\operatorname{softmax}(G_s(\phi(s)))$. A learned mixing coefficient $\beta$ combines the two distributions as $\alpha=\beta\alpha_L+(1-\beta)\alpha_s$. Each branch transforms every lighting token as $e_{k,t}=E_k(L_t)$, and PAMoE returns the augmented feature $\hat L_t=\sum_{k=1}^{4}\alpha_k e_{k,t}$.
\subsection{Adaptive Stage Lighting Retrieval}
Now that we have addressed the core aesthetic challenges through our proposed modules, we integrate them into the unified system as illustrated in Fig.~\ref{fig:pipeline}.
During training, temporally matched music windows and cue-sequence segments are embedded jointly, and professional preference/style annotations supervise the PAMoE conditioning path.
% After that, the pair of data are aligned to be brought into LAMP for representative learning, leading to a pretrained model for inference. The preferences are encoded together with a cue sequence to generate more meaningful representations in the PAMoE module. 
The aligned pairs are fed to LAMP for joint representation learning, while PAMoE injects the dense preference condition into expert-weighted lighting representations. At inference, music and the selected preference query the learned cue space rather than an unconstrained free-form generator.
During inference, the LAMP encoder extracts high-level audio representations from an input music clip.
The system ranks the Cue Sequence Corpus using the cosine similarity $\operatorname{sim}(M_i,L_j)$ defined above and selects the top-$K$ sequences.
The cardinality $K$ is dynamically determined by the Metadata Predictor.
All $K$ retrieved cue sequences are conflict-checked, retargeted to the destination stage, and dispatched to the lighting console for execution.
The sequences are mutually distinct, each monitoring a specific spatial component of the stage illumination.
The Lighting Encoder adopts a transformer-based~\cite{transformer} architecture with four MHA heads, and the Music Encoder is composed of a VGGish network~\cite{vggish} with multilayer perceptrons, all selected based on ablation studies in Section~\ref{sec:ablation}.
%The Metadata Predictor integrates Temporal Convolutional Network density‐count module and Softplus activation to enforce non-negativity.

\section{Dataset: Musilux}
\label{sec:dataset}
\begin{figure*}[ht]
  \centering
  \includegraphics[width=0.95\linewidth]{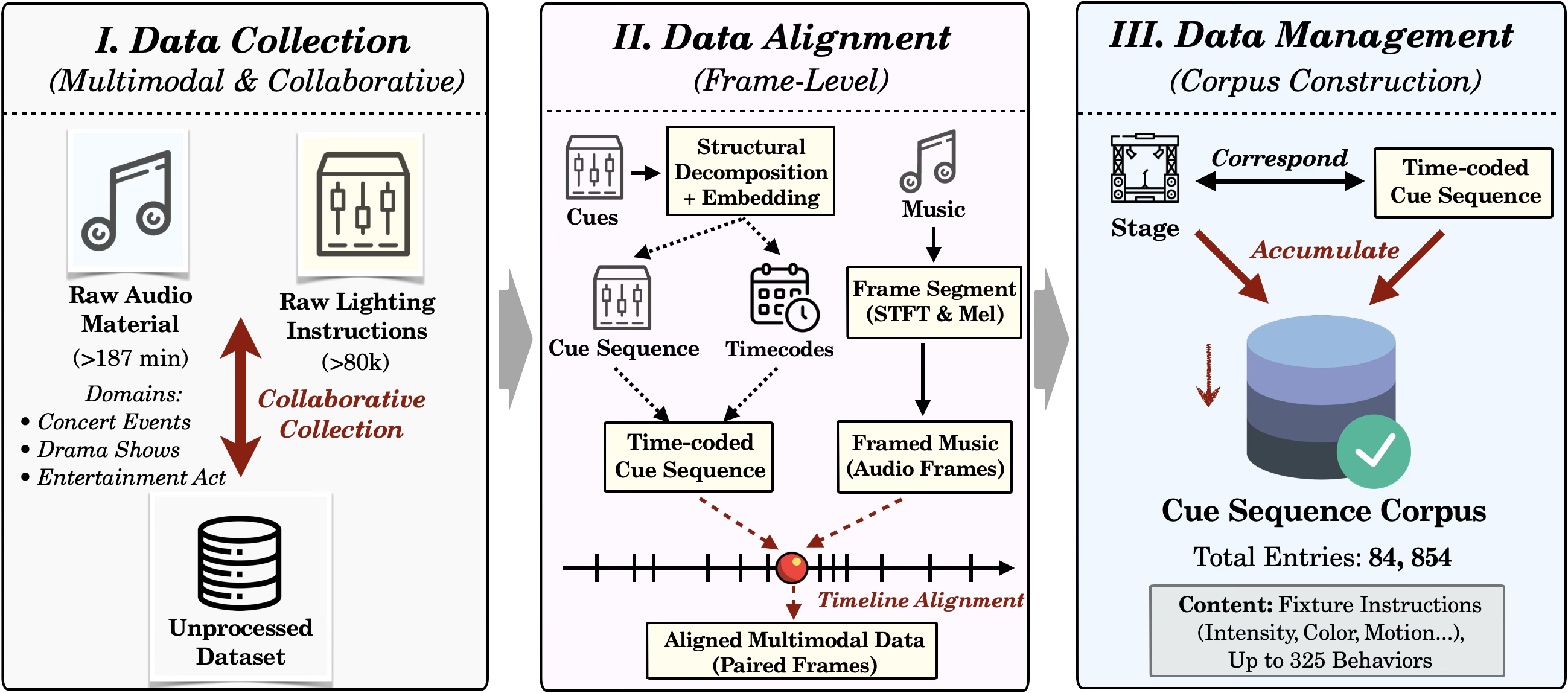}
  \caption{\textbf{Musilux Construction Procedure.} Music and lighting cues are collected, aligned both within cue structures and across modalities, and then aggregated into the Cue Sequence Corpus that constitutes the dataset.}
  \Description{Flow diagram of the Musilux dataset construction procedure: collection, intra-cue alignment, cross-modal alignment, and aggregation into the Cue Sequence Corpus.}
  \label{fig:data_procedure}
\end{figure*}

We construct a multimodal dataset, \Dataset, comprising music and lighting data. We collect Musilux collaboratively with leading institutions in stage lighting design, following the comprehensive construction procedure illustrated in Fig.~\ref{fig:data_procedure}.
Musilux comprises 187 minutes and 22 seconds of musical audio material, accompanied by more than 80,000 associated lighting instructions.
% All audio tracks in the dataset are sourced from royalty-free libraries under permissive licenses. 

\paragraph{Data Collection}
\label{sec:data_collection}
The dataset captures a diverse range of real-world performance scenarios, categorized into the following three primary domains:
\begin{itemize}
  \item \textbf{Concert.} Live performances in which lighting dynamically adapts to musical structure. This domain totals 4,666 seconds and includes Pop, Rock, Dance, Rap, and Electronic subtypes.
  \item \textbf{Drama.} Theatrical productions that employ lighting to enhance narrative and emotional engagement, totaling 3,307 seconds.
  \item \textbf{Entertainment.} Broadcasts, promotional events, and multimedia presentations in which lighting creates a specific atmosphere, totaling 3,269 seconds.
\end{itemize}

Musilux includes 13 distinct stages, each with multiple lighting shows (46.1\% with one program, 30.7\% with two, and 23.2\% with more). This process provides a rich foundation for learning the underlying relationships between modalities.

\paragraph{Data Alignment}
\label{sec:align}
% This process ensures that Musilux is at the unified frame level where each audio frame is paired with its corresponding lighting instructions for coherent multimodal training.
% We first align the Cue Sequence with their corresponding Timecodes annotations so that Each cue is assigned an exact timestamp relative to the musical timeline.
% Subsequently, we perform frame-wise alignment between the music and the time-stamped Cue Sequence. To extract meaningful representations from audio, each frame of music is passed through a Short-Time Fourier Transform and projected onto the Mel scale to approximate the frequency sensitivity of the human auditory system, serving as the foundational input for downstream audio feature extraction.
% In parallel, from each Cue Sequence entry, we extract its essential components: the timestamp and corresponding CueData attributes. We map their timestamps onto the same temporal grid of the audio. After that, the symbolic attributes are embedded into continuous tensor representations using a deliberately designed attribute map which converts discrete lighting attributes into numerical embeddings. This is tested to be the best embedding strategy in Section \ref{sec:embedding_ablation}.
We unify Musilux at frame level by aligning each audio frame with its corresponding lighting instructions.
Cue Sequences are first matched to annotated Timecodes for temporal information.
We then perform frame-wise alignment between the music and time-stamped cues. Audio frames are transformed via Short-Time Fourier Transform (STFT) and projected onto the Mel scale, approximating human auditory perception.
Concurrently, we parse each Cue Sequence into timestamps and attributes. Timestamps are mapped to the frame grid, while lighting attributes are embedded into continuous tensors via a dedicated mapping network.
%This method, shown to outperform other encoding styles in Section~\ref{sec:ablation}, is adopted as our default embedding scheme.

\paragraph{Data Management}
\label{sec:corpus}
As lighting design is formulated as a retrieval-and-adaptation task, we accordingly construct \Corpus of 84,854 entries, each holding multiple CueData units that specify fixture-level instructions (intensity, color, motion). It covers 325 distinct fixture behaviors---color transitions, spatial movements, and speed modulations---forming a behavior-rich foundation.

\section{Experiments}

\subsection{Experiment Setups}
\label{sec:setup}

\paragraph{Configurations}
All experiments were conducted on an NVIDIA GeForce RTX 4090 GPU.
The Metadata Predictor was trained with Adam~\cite{adam} at a learning rate of 0.001, using a total window length of $W=64$ frames and stride $S=4$, achieving convergence within 100 epochs with an early-stopping patience of 8.
LAMP was optimized for 500 epochs with learning rate $5\times10^{-4}$ and weight decay $10^{-4}$; PAMoE was trained jointly on the aligned batches and professional preference labels. Supplementary Section~E reports structural and hyperparameter ablations.
% The metadata predictor is trained with the Adam optimizer at a learning rate of 0.001 for 100 epochs. We employ a contrastive loss with a temperature parameter set to 0.07 when training LAMP. Optimization was performed using AdamW with a learning rate of 5e-4 and a weight decay of 1e-4 over 500 epochs.
Joint LAMP+PAMoE training takes about 1 hour; the Metadata Predictor takes 10--20 minutes; and the complete training pipeline takes approximately 1.0--1.5 hours.
To assess performance across diverse scenarios, we use the disjoint validation/test protocol: validation provides manual references for Table~\ref{tab:test_results}, whereas the test set uses unseen music on an unseen stage.
Evaluation covers virtual and physical environments. We render virtual lighting with GrandMA2 and MA3D, then validate on a physical console and stage.

\begin{figure*}[htbp]
  \centering
  \includegraphics[width=\textwidth]{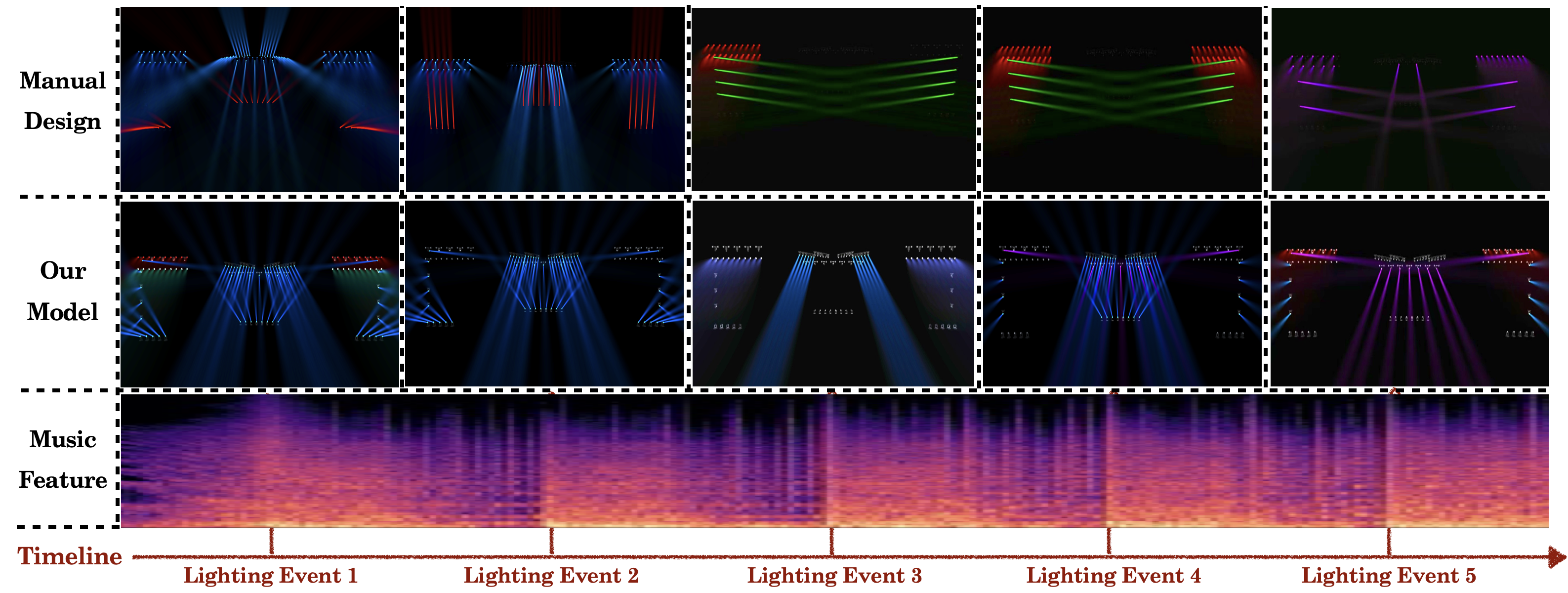}
  \caption{\textbf{Qualitative Results.} AuraLuxMuse and the manual design exhibit different spatial and temporal choices in the same Drama scenario; these examples illustrate scenario-dependent competitiveness rather than universal superiority.}
  \Description{Side-by-side qualitative comparison frames between manually designed lighting and AuraLuxMuse-generated lighting in a Drama scenario, illustrating temporal continuity.}
  \label{fig:qualitative}
\end{figure*}

\paragraph{Implementation}
% When executing a single cue, the stage lighting control software issues DMX signals to fixtures.
%For instance, a signal of Attribute and Value:  \textit{(\text{RGB}, 255), (\text{DIM}, 0.8), (\text{PAN}, 45), (\text{TILT}, 10)} corresponds to emitting red light at 80\% brightness with a specified pan and tilt orientation.
We incorporate a flexible patching mechanism that maps stage lighting between setups.
% For unknown stages, experts only need to assign fixtures, thus enabling generated cue sequences to be seamlessly adapted, reducing the overall design time from days to hours compared to constructing everything manually from scratch.
It begins by accepting the retrieved and adapted cue $\mathbf{C}_{out}=\langle\mathbf{p}_{norm},\mathbf{c}_{req},Attribute,val\rangle$ and a digital twin of the physical stage. The digital-physical address mapping defines a patch table for DMX fixture control.
\footnote{$\mathbf{p}_{norm}=(x,y,z)$ is the spatial fixture address; $\mathbf{c}_{req}=[c_{pan},c_{tilt},c_{rgb},c_{cmy},c_{gobo},\dots]$ encodes required hardware capabilities.}
The assignment cost combines normalized spatial distance with hardware and attribute mismatch; the Kuhn--Munkres solution maps commands to compatible fixtures before console dispatch.
For different stage setups, the mechanism formulates lighting retargeting as a Linear Sum Assignment Problem, projects fixture coordinates into a normalized space, and builds a cost matrix from spatial distances and hardware mismatches. Solving the matrix with the Kuhn--Munkres algorithm yields the optimal permutation that aligns commands with the most appropriate devices, thereby aggregating fixture-level commands into stage-level transitions.
The procedure also resolves conflicts in which two or more retrieved commands assign different values to the same fixture--attribute pair within one execution window.
Supplementary Sections~D, F, and G provide the patching analysis, implementation procedure, console, and fixture details.

% To evaluate the performance of AuraLuxMuse in a diverse range of scenarios, we curate a representative testing set besides the training set. Specifically, one exemplar is selected for each of the three scenario types defined in Section~\ref{sec:data_collection}.
% To evaluate the lighting results in both simulated and physical environments, we conduct a dual-mode visualization. Virtual renderings are produced using the industry-standard grandMA2 and MA3D software suites provided by MA Lighting. For real-world validation, we implement the generated lighting cues in a physical stage setup. Representative static lighting effects are presented in Section~\ref{sec:qualitative}, while complete dynamic results are included in the supplementary video material.

%Representative static effects are shown in Section~\ref{sec:qualitative}
%, while full dynamic visualizations are provided in the supplementary video.

\subsection{Evaluation Metrics}
% As stage lighting constitutes a novel task without standard metrics, we propose customized objective and subjective evaluations.
\paragraph{Objective: FID} 
We use Fr\'echet Inception Distance (FID)~\cite{FID} as a feature-space diagnostic of the distributional discrepancy between retrieved/adapted and reference cue-sequence representations. Because multiple plausible lighting designs can match the same music, FID is not treated as a definitive perceptual or aesthetic metric; lower values indicate closer distributions only in the selected encoder space, not necessarily better perceived quality.

\begin{equation}
\text{FID} = \left\| \mu_r - \mu_g \right\|^2 + \mathrm{Tr} \left( \Sigma_r + \Sigma_g - 2\left( \Sigma_r \Sigma_g \right)^{1/2} \right).
\end{equation}
\noindent $\mu_r$ and $\Sigma_r$ are the empirical mean and covariance of reference cue representations in the lighting-encoder feature space, while $\mu_g$ and $\Sigma_g$ denote AuraLuxMuse outputs in that space.
\paragraph{Subjective: EGSS} 
%To understand how the quality of stage lighting sequences is assessed in practice, we conduct structured interviews with professional lighting designers about the dimensions and standards of evaluation. Through this collaborative process, we identify five key evaluation dimensions for expert-level evaluation, each dimension rated on a scale of 1 to 5, as summarized in Table~\ref{tab:evaluation_expert}, also outlining the criteria used by experts to determine each score.
%, providing a grounded explanation for a subjective but consistent evaluation.
To assess perceptual quality, we interviewed professional lighting designers and identified five expert-rated criteria, each scored on a 5-point Likert scale. We also collected general-audience feedback because readily observable impressions complement expert assessments.

\textbf{We propose the Expert-Guided Subjective Score (\textit{EGSS})}, which combines expert and non-expert perspectives through the convex combination in Equation~(\ref{EGSS}). Supplementary Section~B defines the evaluation dimensions and standards.
\begin{equation}
\label{EGSS}
\text{EGSS} = \lambda \cdot S_{\text{exp}} + (1 - \lambda) \cdot S_{\text{non}}, \quad \lambda \in [0,1].
\end{equation}
\noindent Here, $S_{\text{exp}}$ and $S_{\text{non}}$ denote the average scores from experts and non-experts, respectively. We set $\lambda$ to 0.7 to emphasize expertise while retaining audience perception, based on the ablation studies.
% represent the average scores of expert designers and the general audience, respectively. The scalar  controls the weight of the expert in the final score. We set $\lambda = 0.7$ to emphasize domain expertise while retaining audience perception.

\begin{figure*}[hbt]
  \centering
  \includegraphics[width=\textwidth]{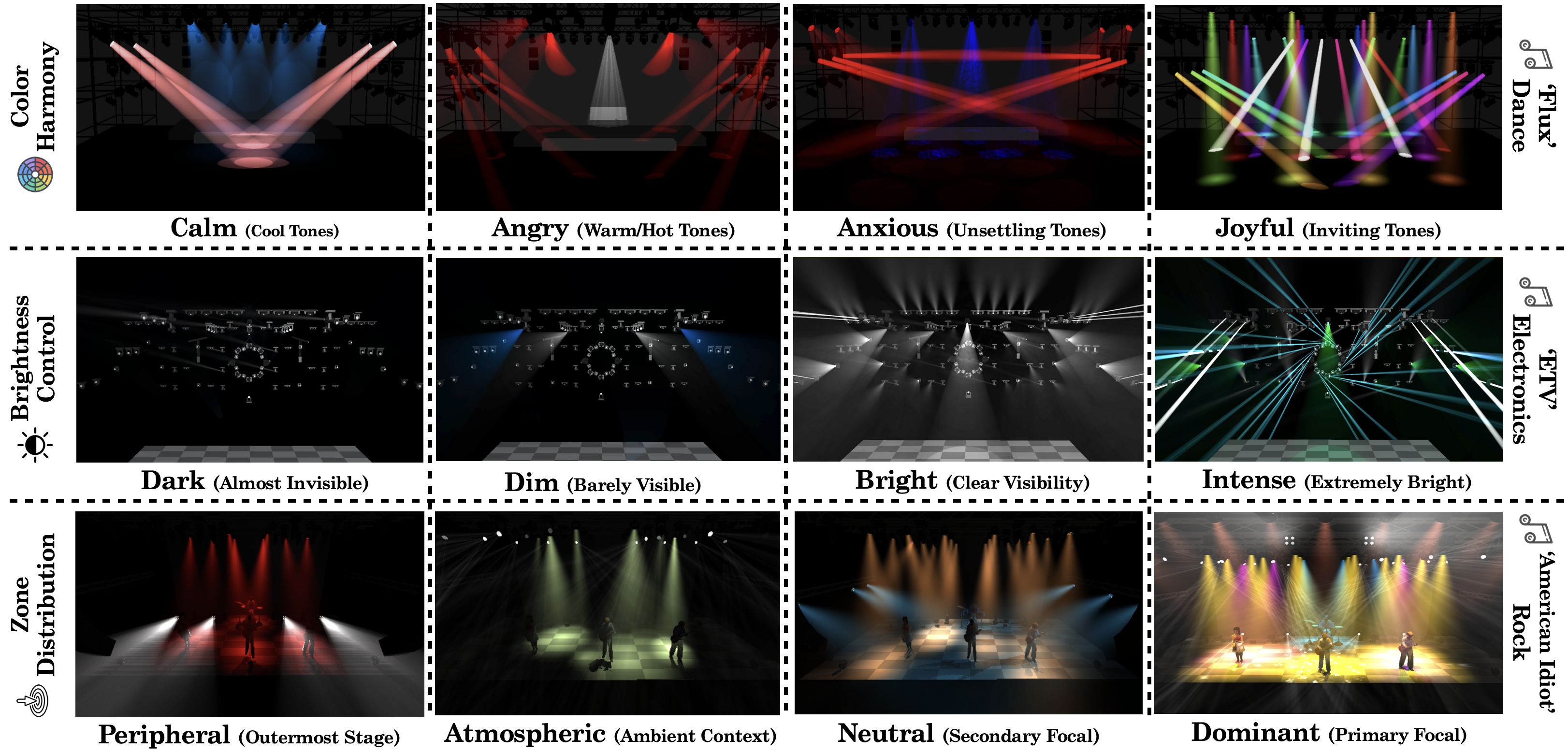}
  \caption{\textbf{Stage Demonstrations across Lighting Styles.} Lighting configurations are visualized across varying levels of Color harmony, Brightness control, and Zone distribution. Definitions of stage-lighting terms are provided in parentheses.}
  \Description{Grid of stage demonstrations showing different combinations of color harmony, brightness levels, and zone distributions for the lighting style space.}
  \label{fig:stages}
\end{figure*}

\subsection{Human Evaluation}
We conduct a user study comparing AuraLuxMuse with manually designed stage lighting across three scenarios. Forty expert participants and 40 general-audience participants rate full-length performance videos presented in randomized, anonymized order. Expert participants are selected based on specialized education and experience in large-scale productions. The expert group comprises the following two subgroups:
\begin{itemize}
  \item \textbf{Student group (15\%).} Participants are 21--24 years old, undergraduates and graduates with practical stage-lighting experience.
  \item \textbf{Teacher group (85\%).} Participants are 35--45 years old, serve as stage-lighting design faculty, and hold advanced academic degrees related to lighting design and production.
\end{itemize}

The two groups exhibit different preferences. Defining the mean difference as AuraLuxMuse minus manual design, experts favor AuraLuxMuse in Concert ($+1.075$, $p=0.00071$) and Entertainment ($+0.748$, $p=0.025$), while the Drama difference is not significant ($-0.366$, $p=0.260$). The expert-only cross-scenario average is a positive but non-significant trend ($+0.486$, $p=0.079$). Non-expert ratings favor manual design on average ($-0.431$, $p=0.0049$). Supplementary Section~B provides the per-scenario breakdown.

\subsection{Experiment Results}
\paragraph{Qualitative Analysis}

\label{sec:qualitative}

For qualitative assessment, we fix representative music frames and generate the corresponding stage lighting. As shown in Fig.~\ref{fig:qualitative}, AuraLuxMuse closely matches manual cues in spatial composition and color dynamics.
At certain timestamps, it demonstrates \textbf{greater visual complexity and temporal continuity}.
We further present three model variations in Supplementary Section~A.
% By integrating expert preferences, PAMoE dynamically synthesizes diverse stylistic variations, making boundless lighting style control at designer's fingertips and facilitating rapid inspirational exploration for professional designing workflows.
% Besides that, we demonstrate the stage of different lighting styles of color, brightness and zone in Fig.~\ref{fig:stages}. It not only shows the difference and the diverse specialization of artistic styles, but also gives out a visualization of the standard and types on basic lighting dimensions.
Representative stages across different stylistic dimensions under various music styles are presented in Fig.~\ref{fig:stages}.
The visualization illustrates preference-conditioned variation across artistic dimensions and music styles, making the abstract lighting dimensions visually interpretable.
% \textcolor{black}{\textbf{(Original: The visualization not only highlights our model's generation ability across artistic dimensions from multiple musical styles but also transforms abstract lighting dimensions into visually intuitive percepts.)}}
% Additional visualizations and full video demonstrations are available in the supplementary materials.
%AuraLuxMuse makes different kinds of stage lighting at the designer's fingertips, serving an inspirational purpose for their artistic ideas.

\paragraph{Quantitative Analysis}
% The model achieves an average inference time of \(7.65\pm0.98\) seconds per input and a throughput of \(3.80\pm0.48\) cues per second in terms of computational performance.
\begin{table}[ht]
\centering
\caption{\textbf{Quantitative Results.} Average expert and non-expert ratings are reported for all scenarios. AuraLuxMuse achieves scenario-dependent competitiveness with the manual counterparts. Values in parentheses are standard deviations. FID measures the representational discrepancy between AuraLuxMuse and the corresponding reference.}
\label{tab:test_results}
\setlength{\tabcolsep}{3pt}
\resizebox{\columnwidth}{!}{%
\begin{tabular}{@{}cccccc@{}}
\toprule
\textbf{Scenario} & \textbf{Source} & \textbf{FID} & $\mathbf{S_{exp}}$ $\uparrow$ & $\mathbf{S_{non}}$  & \textbf{EGSS} $\uparrow$ \\
\midrule
\multirow{2}{*}{\textbf{Concert}}
    & Manual & \multirow{2}{*}{8.21} & 2.58 {\scriptsize (1.33)} & 4.31 {\scriptsize (0.73)} & 3.09 {\scriptsize (1.15)} \\
    & AuraLuxMuse & & 3.65 {\scriptsize (1.00)} & 3.69 {\scriptsize (0.88)} & 3.66 {\scriptsize (0.75)} \\
\addlinespace[3pt]
\multirow{2}{*}{\textbf{Drama}}
    & Manual & \multirow{2}{*}{4.47} & 3.08 {\scriptsize (1.10)} & 4.03 {\scriptsize (0.93)} & 3.36 {\scriptsize (1.04)}\\
    & AuraLuxMuse & & 2.71 {\scriptsize (1.24)} & 3.27 {\scriptsize (1.11)}  & 2.88 {\scriptsize (0.93)} \\
\addlinespace[3pt]
\multirow{2}{*}{\textbf{Entertainment}}
    & Manual & \multirow{2}{*}{2.00} & 2.46 {\scriptsize (1.25)} & 3.31 {\scriptsize (0.98)} & 2.71 {\scriptsize (1.16)} \\
    & AuraLuxMuse & & 3.21 {\scriptsize (1.25)} & 3.39 {\scriptsize (1.13)} & 3.26 {\scriptsize (0.94)} \\
\midrule
\multirow{2}{*}{\textbf{Average}}
    & Manual & \multirow{2}{*}{4.89} & 2.71 {\scriptsize (1.23)} & \textbf{3.88} {\scriptsize (0.88)} & 3.05 {\scriptsize (1.12)} \\
    & AuraLuxMuse & & \textbf{3.19} {\scriptsize (1.17)} & 3.45 {\scriptsize (1.05)} & \textbf{3.27} {\scriptsize (0.88)}\\
\bottomrule
\end{tabular}%
}
\end{table}

The model achieves an average inference time of \({\bm{7.65 \pm 0.98}}\)\ \textbf{seconds}.
Manual designs and AuraLuxMuse outputs are compared across all scenarios in Table~\ref{tab:test_results}.
Across the evaluated scenarios, the expert-only comparison is scenario dependent: AuraLuxMuse is favored in Concert and Entertainment, while the Drama difference and the cross-scenario expert average are not statistically significant. These results indicate that the system is competitive with manual designs in the evaluated scenarios while operating with substantially greater efficiency.
In addition to comparisons with manual designs, we compare the similarity-based retrieval strategy with several alternative retrieval methods using both standard retrieval metrics and the real-world performance metric EGSS.
As shown in Table~\ref{tab:strategy}, similarity retrieval performs best among the tested retrieval strategies. The rule-based baseline partitions the cue corpus using coarse audio heuristics before sampling valid professional cues. Random retrieval is a nontrivial corpus-prior lower bound because it also inherits metadata-predicted timing and samples console-valid cues. However, it lacks music-conditioned semantic alignment.
%we note that no prior deeep-learning methods applied to stage lighting, resulting in no existing methodological baselines in the domain.
% Table~\ref{tab:test_results} compares manually designed stage lighting with results generated by AuraLuxMuse across all scenario types. On average, the AuraLuxMuse output achieves comparable subjective scores from both experts and the general audience. Notably, in certain scenarios, expert evaluations indicate that AuraLuxMuse occasionally surpasses manual designs. These results suggest that the system produces high-quality stage lighting effects that are often indistinguishable from human-designed ones.
Furthermore, we assess subjective performance across various music styles. We evaluate the model on styles present in Musilux to demonstrate its ability to reproduce known musical genres and test generalization on styles not included in the training set.
Detailed results of these evaluations are presented in Table~\ref{tab:music_styles}.
\begin{table}[ht]
  \caption{\textbf{Comparison of Retrieval Strategies.} We compare similarity retrieval with random, position-based, frequency-based, nearest-neighbor, and rule-based strategies. Values in parentheses are standard deviations.}
  \label{tab:strategy}
\centering
\footnotesize
\resizebox{\columnwidth}{!}{%

\begin{tabular}{lcccc}
\toprule
\textbf{Strategy} & \textbf{FID} $\downarrow$ & \textbf{Precision@5} $\uparrow$ & \textbf{mAP} $\uparrow$ & \textbf{EGSS} $\uparrow$ \\
\midrule
Random & 6.32 & 0.07 & 0.03 & 2.71 {\scriptsize (0.72)} \\
Position-based & 25.14 & 0.07 & 0.04 & 1.98 {\scriptsize (0.73)} \\
Frequency-based & 16.32 & 0.07 & 0.03 & 2.44 {\scriptsize (0.76)} \\
Nearest-neighbor & 21.65 & 0.08 & 0.04 & 2.10 {\scriptsize (0.73)} \\
Rule-based & 6.61 & 0.06 & 0.02 & 2.81 {\scriptsize (0.81)} \\
\midrule
\textbf{Similarity-based} & \textbf{4.89} & \textbf{0.17} & \textbf{0.10} & \textbf{3.27} {\scriptsize \textbf{(0.88)}} \\
\bottomrule
\end{tabular}
}
\end{table}

\paragraph{Ablation Studies}

\aptLtoX{\begin{table}[t]
  \centering
\caption{\textbf{Subjective Results across Music Styles.} EGSS is reported for six validation styles and six test styles. Validation uses held-out examples with manual references; test tracks and the test stage are excluded from training.}
  \label{tab:music_styles}
\resizebox{\columnwidth}{!}{%
    \begin{tabular}{ccccccc}

    \multicolumn{7}{c}{\textbf{(a) Validation Music Styles}}\\
    \hline
      \textbf{Music Styles} & Pop & Rock & Opera & Dance & Rap & Electronic \\
      \hline
      \textbf{EGSS}  & 3.02 {\scriptsize (1.23)}   & 3.66 {\scriptsize (0.75)}   & 2.88 {\scriptsize (0.93)}   & 3.26 {\scriptsize (0.94)}   & 2.83 {\scriptsize (1.04)}   & 2.84 {\scriptsize (1.31)}        \\
      \hline
    \end{tabular}}
    \centering
\resizebox{\columnwidth}{!}{%
    \begin{tabular}{ccccccc}
\multicolumn{7}{c}{\textbf{(b) Test Music Styles}}\\
      \hline
      \textbf{Music Styles} & Funk & Disco & Jazz & Country & Folk & Classical \\
      \hline
      \textbf{EGSS}  &   3.25 {\scriptsize (1.17)}  & 2.94 {\scriptsize (1.20)}     & 3.51 {\scriptsize (1.38)}    & 2.98 {\scriptsize (1.36)}       & 2.70 {\scriptsize (0.97)}    & 3.05 {\scriptsize (1.31)}          \\
      \hline
    \end{tabular}}
\end{table}}{\begin{table}[t]
  \centering
\caption{\textbf{Subjective Results across Music Styles.} EGSS is reported for six validation styles and six test styles. Validation uses held-out examples with manual references; test tracks and the test stage are excluded from training.}
  \label{tab:music_styles}
  \begin{subtable}[t]{\linewidth}
    \centering
    \caption{\textbf{Validation Music Styles}}
    \resizebox{\linewidth}{!}{%
    \begin{tabular}{ccccccc}
      \toprule
      \textbf{Music Styles} & Pop & Rock & Opera & Dance & Rap & Electronic \\
      \midrule
      \textbf{EGSS}  & 3.02 {\scriptsize (1.23)}   & 3.66 {\scriptsize (0.75)}   & 2.88 {\scriptsize (0.93)}   & 3.26 {\scriptsize (0.94)}   & 2.83 {\scriptsize (1.04)}   & 2.84 {\scriptsize (1.31)}        \\
      \bottomrule
    \end{tabular}%
    }
  \end{subtable}
  \begin{subtable}[t]{\linewidth}
    \centering
    \caption{\textbf{Test Music Styles}}
    \resizebox{\linewidth}{!}{%
    \begin{tabular}{ccccccc}
      \toprule
      \textbf{Music Styles} & Funk & Disco & Jazz & Country & Folk & Classical \\
      \midrule
      \textbf{EGSS}  &   3.25 {\scriptsize (1.17)}  & 2.94 {\scriptsize (1.20)}     & 3.51 {\scriptsize (1.38)}    & 2.98 {\scriptsize (1.36)}       & 2.70 {\scriptsize (0.97)}    & 3.05 {\scriptsize (1.31)}          \\
      \bottomrule
    \end{tabular}%
    }
  \end{subtable}

\end{table}}

\label{sec:ablation}
We ablate the LAMP backbones~\cite{resnet,transformer}, Metadata Predictor, and PAMoE; compare contrastive~\cite{audioclip, CLAP, SLAP} and generative~\cite{bart, autoformer, T5, transformermusic} baselines; analyze $K$, soft/hard-alignment weights $w_{i,p}$, and lighting embeddings; and test metric sensitivity and the FID--EGSS correlation. Supplementary Section~E provides the full designs, quantitative results, training dynamics, and sensitivity analyses.
\subsection{Discussion}
Across the evaluated scenarios, AuraLuxMuse is competitive with manual designs while producing candidate cues with an average inference time of $7.65 \pm 0.98$ seconds under the reported setup. It retrieves and adapts professional cues rather than replacing unconstrained human creativity. This approach outperforms the tested sequence-to-sequence generative baselines because small lighting-parameter perturbations can yield divergent effects and end-to-end models often fail to capture the constraints required for high-fidelity synthesis.

More broadly, our results show that an underexplored modality can be represented and integrated into a computational system, suggesting a path toward multimodal modeling of other unconventional modalities. AuraLuxMuse is, to the best of our knowledge, the first retrieval-and-adaptation system to combine \textbf{high quality, efficiency, and editable flexibility} in automated stage lighting design.

\subsection{Limitations}
AuraLuxMuse is an auxiliary-design system. It retrieves from a fixed expert-authored corpus and therefore cannot create fixture behaviors or artistic structures absent from that corpus. Music without a clear groove, script-only material without an audio timing signal, and rapidly changing styles remain difficult. Results do not establish population-scale generalization to all genres due to the size of validation and test set.
The visual outputs can be over-bright or washed out. Person-following spotlights, narrative focus changes, and other high-level show decisions still require manual authoring and final approval.
FID measures distance only in the selected encoder space and is an imperfect objective metric.
The system supports efficient prototyping and inspection of candidate lighting productions; it does not replace professional judgment or creative work.

\section{Conclusion}
We introduce \textbf{AuraLuxMuse}, a multimodal system for intelligent stage lighting design via preference-conditioned retrieval and retargeting, together with the first professional lighting-music dataset Musilux.
Our approach advances the field through a dedicated representation module that correlates music with lighting, a previously underexplored modality.
PAMoE enables the injection of personalized preferences and the capture of underlying style characteristics.
AuraLuxMuse marks a significant step toward AI-assisted stagecraft, offering both theoretical contributions to multimodal learning and practical value in real-world applications.

% We introduce AuraLuxMuse, a multi-modal and preference-adaptive system for intelligent stage lighting design that extends beyond naive multi-modality alignment. 
% Through a combination of soft and hard alignment strategies and a dedicated embedding system of LAMP, our method captures nuanced temporal and semantic correspondences between music and stage lighting, which is a modality that has never been discussed or parameterized before.
% Central to our architecture is the PAMoE module, which not only enables the injection of personalized stylistic preferences during inference but also ensures a more profound representation of the stage lighting capturing the underlying intricate style information.

% Empirically, AuraLuxMuse achieves performance on par with human-designed lighting in both quantitative metrics and perceptual evaluations, with many outputs perceived as indistinguishable from professional productions. The system operates at unprecedented speed compared to manual design, producing full-scene lighting for entire performances within approximately 10 seconds, representing a substantial improvement in production efficiency. To the best of our knowledge, AuraLuxMuse is the first system in the domain to simultaneously realize high perceptual fidelity, user-controllable generation, and practical runtime performance, marking a significant step toward AI-collaborative stagecraft.

\begin{acks}
This work is supported by the
\grantsponsor{nsfc-62402306}
{National Natural Science Foundation of China}
{}
(Grant No.~\grantnum{nsfc-62402306}{62402306}), the
\grantsponsor{shanghai-nsf-24ZR1422400}
{Natural Science Foundation of Shanghai}
{}
(Grant No.~\grantnum{shanghai-nsf-24ZR1422400}{24ZR1422400}), and the
\grantsponsor{shuguang-24SG50}
{Shuguang Program of Shanghai Education Development Foundation and Shanghai Municipal Education Commission}
{}
(Project No.~\grantnum{shuguang-24SG50}{24SG50}).
This work is also supported by the \grantsponsor{shanghai-ai-performing-arts}
{Shanghai Universities Industry-Education-Research Collaborative Education Team}
{}, specifically the ``AI + Performing Arts'' Education Team.
\end{acks}

\bibliographystyle{ACM-Reference-Format}
\bibliography{reference}

\clearpage
\twocolumn[{%
  \centering
  {\Large\bfseries Supplementary Material\par}
  \vspace{0.4em}
  {\normalsize for AuraLuxMuse: Adaptive Fusion Modeling for Aesthetic Stage Lighting Design with Music and Expert Guidance\par}
  \vspace{1em}
}]
\appendix
\suppressfloats[t]
\setcounter{figure}{0}
\setcounter{table}{0}
\renewcommand{\thefigure}{S\arabic{figure}}
\renewcommand{\thetable}{S\arabic{table}}
\appendix
\hypersetup{linkcolor=black, linktoc=all}
\phantomsection
%\addcontentsline{toc}{section}{Appendix of AuraLuxMuse}
\noindent{\large
\begin{itemize}
    \item \hyperref[A]{Section A: Cross-Scenario Visualization of Lighting Effects}
    \item \hyperref[B]{Section B: Evaluation Dimensions and Standards}
    \item \hyperref[C]{Section C: Preference and Style Details}
    \item \hyperref[D]{Section D: Patching Mechanism Analysis}
    \item \hyperref[E]{Section E: Ablation Studies}
    \item \hyperref[F]{Section F: Stage Lighting Environments}
    \item \hyperref[G]{Section G: Stage Lighting Implementation Procedure}
    \item \hyperref[H]{Section H: Dataset Split and Generalization Analysis} 
\end{itemize}
}

\section{Cross-Scenario Visualization of Lighting Effects}
\label{A}

\begin{figure*}[!t]
  \centering
  \includegraphics[width=0.885\linewidth]{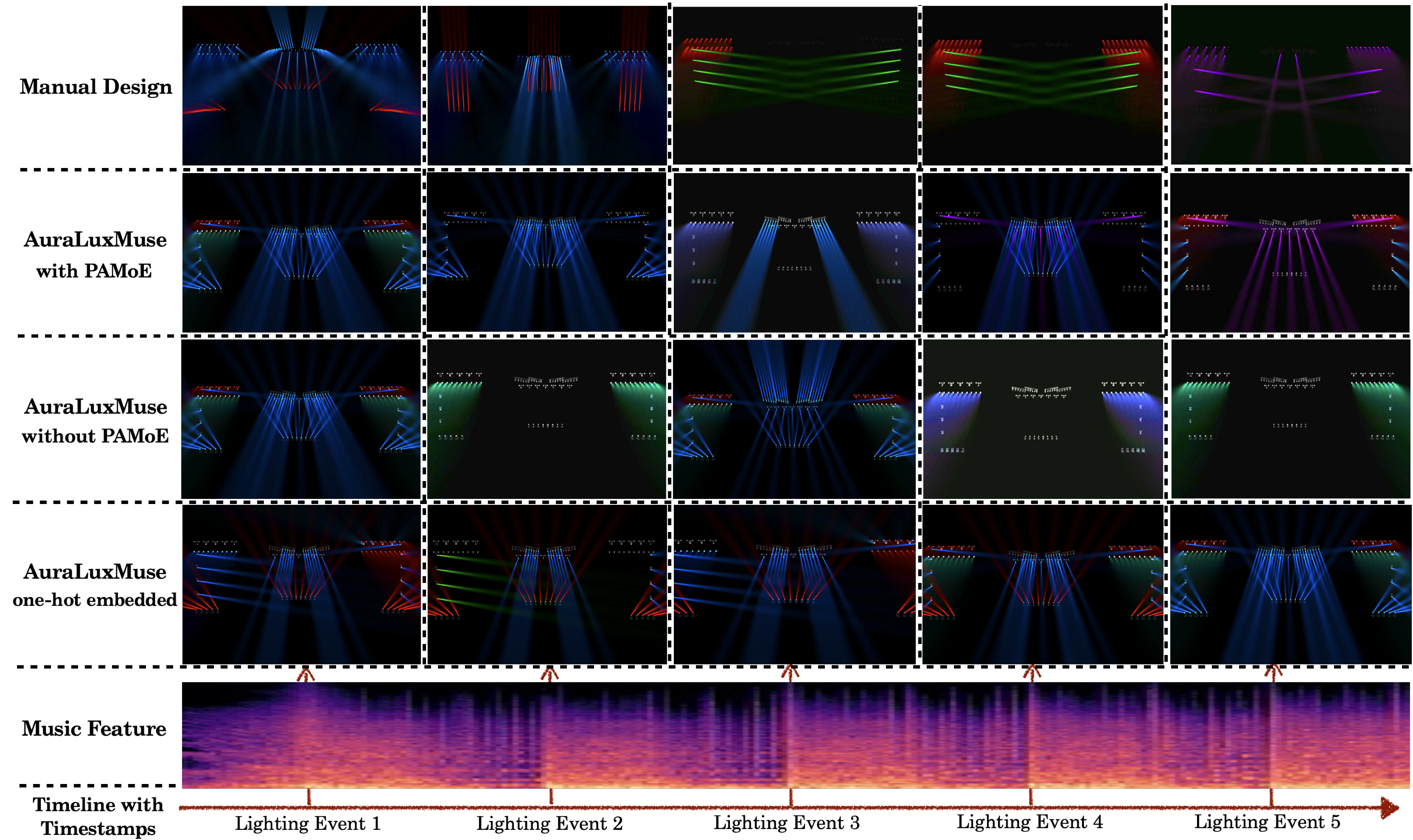}
  \caption{\textbf{Qualitative Results for AuraLuxMuse Variants in the Drama Scenario.} Three architecture variants produce distinct retrieved/adapted lighting outputs as editable references for professional designers.}
  \Description{A grid of stage-lighting frames comparing manual design with three AuraLuxMuse variants in the Drama scenario.}
  \label{fig:full}
\end{figure*}

\begin{figure*}[!t]
  \centering
  \includegraphics[width=\linewidth]{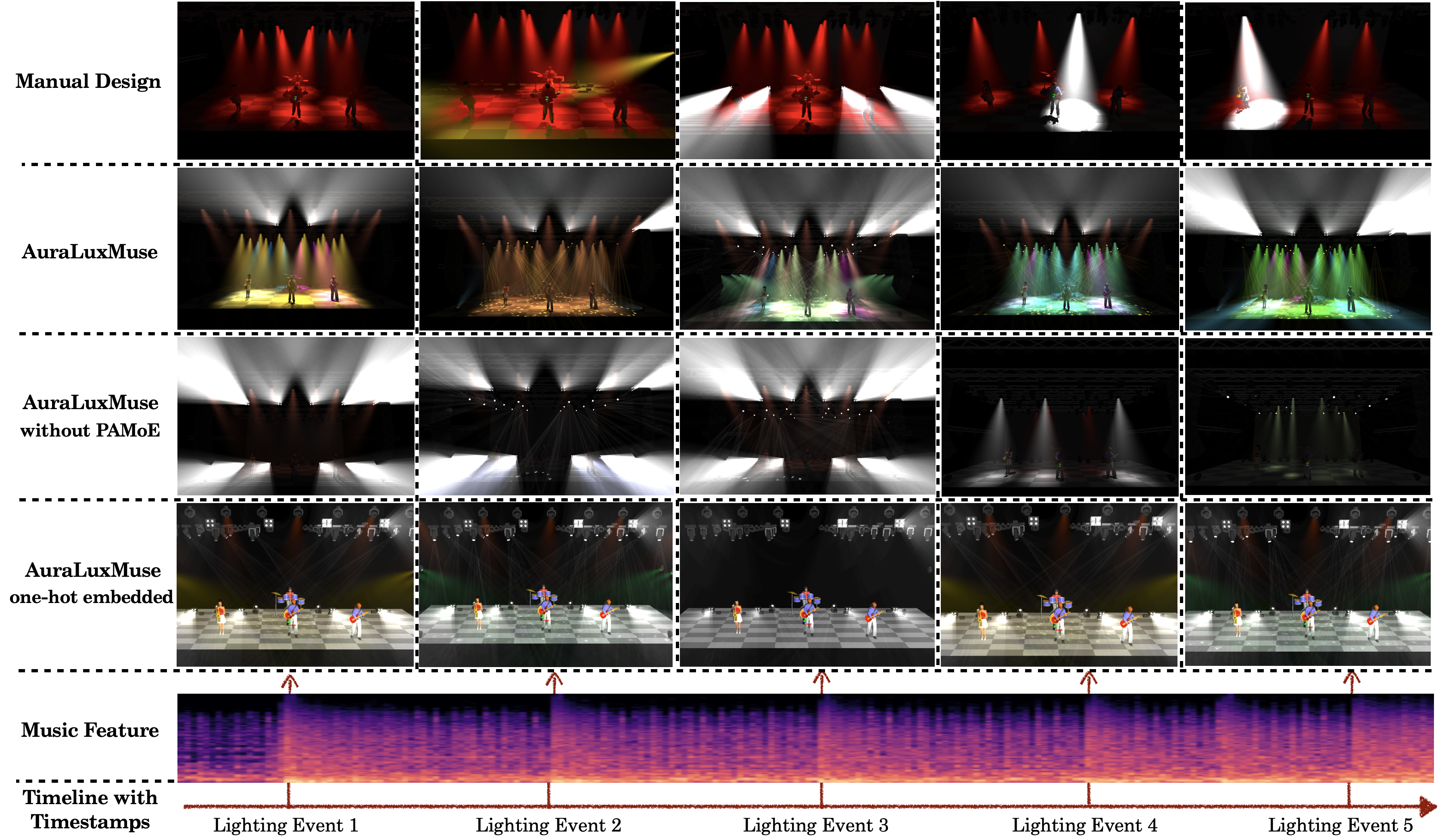}
  \caption{\textbf{Qualitative Results in the Concert Scenario.} Consecutive frames compare manual lighting with AuraLuxMuse retrieved/adapted outputs under the same scenario; the variants show how the model structure changes cue selection. }
  \Description{A grid of stage-lighting frames comparing manual design with AuraLuxMuse variants in the Concert scenario, together with music features and event timestamps.}
  \label{fig:concert}
\end{figure*}

\begin{figure*}[!htbp]
    \centering
  \includegraphics[width=\linewidth]{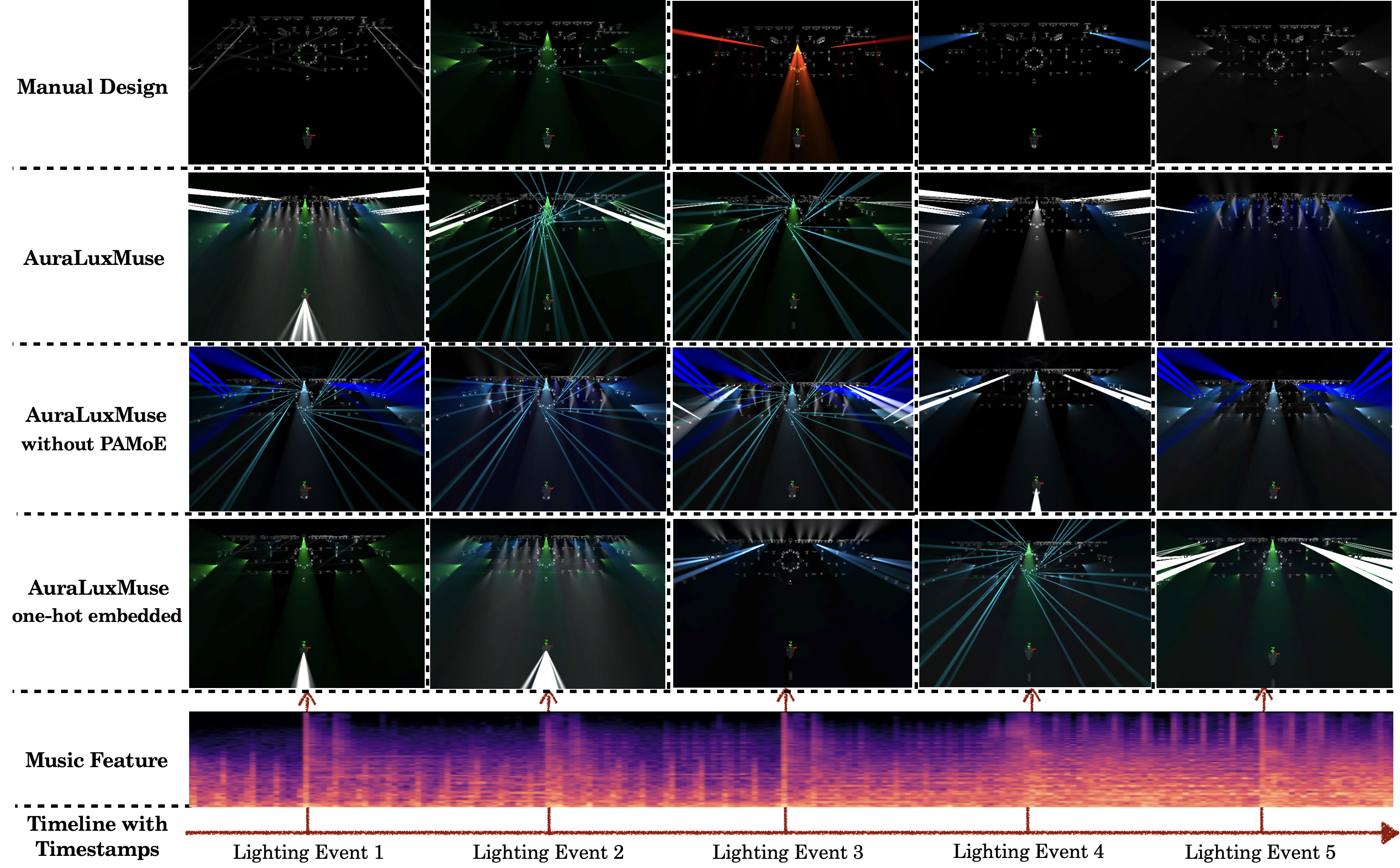}
  \caption{\textbf{Qualitative Results in the Entertainment Scenario.} Consecutive frames compare a manual design with AuraLuxMuse retrieved/adapted cues under the same Entertainment scenario. The example illustrates different choices in visual complexity and atmosphere without implying universal superiority. }
  \Description{A grid of consecutive Entertainment-scenario lighting frames comparing manual design with AuraLuxMuse retrieved and adapted cues, plus music features and event timestamps.}
  \label{fig:entertainment}
\end{figure*}

To further illustrate the system, we present qualitative visualizations across multiple performance scenarios. Figures~\ref{fig:full}, \ref{fig:concert}, and \ref{fig:entertainment} compare manual designs with AuraLuxMuse outputs for the three scenarios. These examples show music-responsive cue choices and scenario-dependent strengths, but they do not establish universal superiority over professional designers. Instead, AuraLuxMuse provides an editable retrieval-and-adaptation workflow that substantially reduces operating time. 

In the illustrated Concert example, AuraLuxMuse selects cues that increase on-stage visibility for instrumentalists and differ stylistically from the manual design. In the Entertainment example, the retrieved cues emphasize vibrant and complex behavior. These are qualitative observations for the shown cases, not general claims about audience engagement or superiority. 

\section{Evaluation Dimensions and Standards}
\label{B}

\label{sec:evaluation}

\subsection{Experts}
\label{B1}

\begin{table}[t]
  \caption{\textbf{Lighting Evaluation for Experts: Color Harmony.} This dimension rates the aesthetic balance and blending of stage-lighting colors across the complete visual composition of each scene.}
  \label{tab:color_harmony}
  \centering
  \begin{tabular}{cc}
    \toprule
    \textbf{Ratings} & \textbf{Color Harmony} \\
    \midrule
    \textbf{1} & Harsh, unbalanced colors; poor harmony \\
    \textbf{2} & Colors clash; mildly unbalanced \\
    \textbf{3} & Colors blend acceptably; moderate harmony \\
    \textbf{4} & Colors blend smoothly; strong harmony \\
    \textbf{5} & Colors blend flawlessly; outstanding harmony \\
    \bottomrule
  \end{tabular}
\end{table}

\begin{table}[t]
  \caption{\textbf{Lighting Evaluation for Experts: Brightness Control.} This dimension rates the consistency and adjustment of brightness across an entire performance sequence under review.}
  \label{tab:brightness_control}
  \centering
  \begin{tabular}{cc}
    \toprule
    \textbf{Ratings} & \textbf{Brightness Control} \\
    \midrule
    \textbf{1} & Uneven brightness; excessively dim/bright \\
    \textbf{2} & Uneven brightness; slightly harsh \\
    \textbf{3} & Reasonable brightness control; minor issues \\
    \textbf{4} & Consistent brightness; well-adjusted \\
    \textbf{5} & Perfect brightness; optimal balance \\
    \bottomrule
  \end{tabular}
\end{table}
\begin{table}[t]
  \caption{\textbf{Lighting Evaluation for Experts: Zone Distribution.} This dimension rates evenness and coverage across the visible stage area.}
  \label{tab:zone_distribution}
  \centering
  \begin{tabular}{cc}
    \toprule
    \textbf{Ratings} & \textbf{Zone Distribution} \\
    \midrule
    \textbf{1} & Uneven zones; inadequate coverage \\
    \textbf{2} & Uneven zones; noticeable gaps \\
    \textbf{3} & Fair zoning; slight inconsistencies \\
    \textbf{4} & Even zoning; effective coverage \\
    \textbf{5} & Flawless zoning; complete coverage \\
    \bottomrule
  \end{tabular}
\end{table}

\begin{table}[t]
  \caption{\textbf{Lighting Evaluation for Experts: Rhythmic Tempo.} This dimension rates lighting synchronization with performance rhythm.}
  \label{tab:rhythmic_tempo}
  \centering
  \begin{tabular}{cc}
    \toprule
    \textbf{Ratings} & \textbf{Rhythmic Tempo} \\
    \midrule
    \textbf{1} & No rhythm; disorganized timing \\
    \textbf{2} & Weak rhythm; moderately off \\
    \textbf{3} & Average rhythm; reasonably aligned \\
    \textbf{4} & Good rhythm; accurately timed \\
    \textbf{5} & Excellent rhythm; impeccably timed \\
    \bottomrule
  \end{tabular}
\end{table}

\begin{table}[ht]
  \caption{\textbf{Lighting Evaluation for Experts: Narrativity.} This dimension rates how effectively lighting conveys plot and emotion.}
  \label{tab:narrativity}
  \centering
  \begin{tabular}{cc}
    \toprule
    \textbf{Ratings} & \textbf{Narrativity} \\
    \midrule
    \textbf{1} & Fails to convey plot / emotion \\
    \textbf{2} & Weakly conveys plot / emotion \\
    \textbf{3} & Moderately conveys plot / emotion \\
    \textbf{4} & Clearly conveys plot / emotion \\
    \textbf{5} & Strongly conveys plot / emotion \\
    \bottomrule
  \end{tabular}
\end{table}

Tables~\ref{tab:color_harmony}--\ref{tab:narrativity} define the five expert evaluation dimensions---color harmony, brightness control, zone distribution, rhythmic tempo, and narrativity---and their rating standards.

Before rating, all professionals reviewed the definitions and standards. Each rated 12 scenes in five dimensions: one manual design and three AuraLuxMuse variants per scenario. Section~\ref{sec:style_standard} details the dimensions used in evaluation.

\subsection{Non-Experts}
\label{B2}
\begin{table*}[ht]
  \caption{\textbf{Lighting Evaluation Dimensions (Relative to Music) for Non-Experts.} The criteria capture spectators' intuitive perceptions of performance quality across the three evaluated scenarios.}
  \label{tab:evaluation_nonexpert}
  \centering
  \small
  \setlength{\tabcolsep}{3pt} 
  
  \begin{tabular}{@{}c*{3}{>{\centering\arraybackslash}p{0.27\linewidth}}@{}}
    \toprule
    % 2. 使用 [-0.8ex] 将 Ratings 垂直向下微调，使其在复合高度的表头中视觉居中
    \multirow{2}{*}[-1ex]{\textbf{Ratings}} & \multicolumn{3}{c}{\textbf{Evaluation Dimensions}} \\
    \cmidrule{2-4} 
    & \makecell[c]{\textbf{Rhythmic} \\ \textbf{Alignment}} & \makecell[c]{\textbf{Thematic} \\ \textbf{Suitability}} & \makecell[c]{\textbf{Visual} \\ \textbf{Consistency}} \\
    \midrule
    \textbf{1} & No alignment & Not suited & Extremely inconsistent \\
    \textbf{2} & Poor alignment & Poorly suited & Somewhat inconsistent \\
    \textbf{3} & Basic alignment & Basically suited & Basically consistent \\
    \textbf{4} & Fair alignment & Fairly suited & Generally consistent \\
    \textbf{5} & Perfect alignment & Perfectly suited & Always consistent \\
    \bottomrule
  \end{tabular}
\end{table*}

For non-expert participants, we designed a clear and accessible protocol covering rhythmic accuracy, thematic coherence, and narrative visual integrity. Table~\ref{tab:evaluation_nonexpert} details the criteria, which general-audience participants perceived as straightforward.
For example, lighting actions synchronized with musical downbeats indicated strong rhythmic alignment. Audience members also assessed whether the lighting conveyed the music's theme and remained stylistically consistent. All participants reviewed the definitions and standards before rating every scene in the overall study.

\subsection{Human-Study Statistical Details }
Values in parentheses in Tables~1--3 of the main paper are standard deviations. Defining each difference as AuraLuxMuse minus manual design, the expert results are: Concert $+1.075$ (95\% CI $[0.480,1.671]$, $p=0.00071$), Drama $-0.366$ ($[-1.017,0.286]$, $p=0.260$), Entertainment $+0.748$ ($[0.098,1.398]$, $p=0.025$), and the cross-scenario expert average $+0.486$ ($[-0.059,1.031]$, $p=0.079$). Non-expert descriptive results are Concert $3.69\pm0.88$ versus $4.31\pm0.73$ (difference $-0.62$), Drama $3.27\pm1.11$ versus $4.03\pm0.93$ (difference $-0.76$), and Entertainment $3.39\pm1.13$ versus $3.31\pm0.98$ (difference $+0.08$); the cross-scenario non-expert difference is $-0.431$ (95\% CI $[-0.723,-0.138]$, $p=0.0049$). EGSS is likewise scenario dependent: AuraLuxMuse is higher in Concert ($3.66\pm0.75$ versus $3.09\pm1.15$, $p=0.010$) and Entertainment ($3.26\pm0.94$ versus $2.71\pm1.16$, $p=0.024$), while manual design is higher in Drama ($2.88\pm0.93$ versus $3.36\pm1.04$, $p=0.044$); the overall EGSS difference is not significant ($p=0.282$). Based on paired ratings, 72.5\% of experts and 40.0\% of non-experts give AuraLuxMuse a higher overall rating. These results motivate the narrower claim of scenario-dependent competitiveness.

\section{Preference and Style Details}
\label{C}
\begin{table}[t]
    \centering
  \caption{\textbf{Overview of Lighting Style Dimensions.} Color, Brightness, Zone, and Rhythm are mapped to representative attributes.}
  \label{tab:lighting_styles}
  \small
  \begin{tabular}{@{}p{0.27\columnwidth}p{0.66\columnwidth}@{}}
    \toprule
    \makecell[c]{\textbf{Lighting Styles}}&\textbf{Attributes}\\
    \midrule
    Color Harmony & \textit{Joyful, Calm, Thrilled, Indifferent, Surprised, Confused, Depressed, Anxious, Angry, Fearful} \\
    Brightness Control & \textit{Dark, Dim, Moderate, Bright, Intense}\\
    Zone Distribution & \textit{Peripheral, Atmospheric, Neutral, Attention, Dominant Zone} \\
    Rhythmic Tempo & \textit{Static, Slow, Moderate, Fast, Rapid }\\
  \bottomrule
  \end{tabular}
\end{table}

\subsection{Style Dimensions Specification}
\label{C1}
\label{sec:style_standard}

This section presents the stage-lighting style dimensions and their corresponding attributes, as summarized in Table~\ref{tab:lighting_styles}. The attribute definitions and assignment standards follow.
Professional lighting designers assign one ordinal value for each of Color, Brightness, Zone, and Rhythm to each labeled source program using the style taxonomy in Table~\ref{tab:lighting_styles} and the detailed level definitions in Tables~\ref{tab:brightness_ranges}, \ref{tab:zone}, and \ref{tab:tempo}. The 33 available labeled programs provide the tuples $s=[s_c,s_b,s_z,s_r]$ used in the gate analysis. Each value is represented by a 32-D embedding, and the four vectors form the 128-D condition $\phi(s)$. During joint training, the tuple accompanies every music--cue window from that program; EGSS is never used as a training label. Inter-rater agreement was not measured for these style annotations, which limits reliability claims based on the labels.

\subsection{Style Details}

\label{C2}
Color harmony is a foundational stylistic dimension in stage-lighting design because it directly shapes the viewer's emotional perception and interpretation of a performance. Rooted in color theory and psychological aesthetics, color harmony describes the visual compatibility of color combinations: how well different hues, saturations, and brightness levels cohere within a lighting composition. We operationalize color harmony by embedding its semantics in a color--emotion mapping space informed by color psychology and cinematographic lighting design.

Brightness control is a critical dimension for evaluating and guiding stage-lighting design, offering both subjective stylistic interpretation and objective physical quantification. Lighting experts possess an intuitive grasp of illuminance ranges that aligns with defined lux intervals used in lighting-design standards. As shown in Table~\ref{tab:brightness_ranges}, Brightness is categorized into five discrete levels associated with distinct perceptual descriptions and lux ranges.
\begin{table}[t]
\centering
\caption{\textbf{Brightness Attribute Levels.} Each level is defined by a perceptual description and illuminance range for preference specification.}
\label{tab:brightness_ranges}
\begin{tabular}{ccc}
\toprule
\textbf{Level} & \textbf{Description} & \textbf{Illuminance (lux)} \\
\midrule
Dark   & Almost invisible        & 0–20     \\
Dim        & Barely visible            & 20–100   \\
Moderate   & Visually comfortable & 100–300  \\
Bright     & Clear visibility         & 300–700  \\
Intense & Extremely bright         & 700+     \\
\bottomrule
\end{tabular}
\end{table}

Zone distribution organizes visual-prominence regions using principles of art and perception. These regions shape audience focus and narrative structure; Table~\ref{tab:zone} defines the attributes.
\begin{table}[t]
  \centering
  \caption{\textbf{Zone Distribution Attribute Levels.} Each level defines a degree of visual prominence within a scene.}
  \label{tab:zone}
  \begin{tabular}{cc}
    \toprule
    \textbf{Level} & \textbf{Description} \\
    \midrule
    Peripheral & Invisible boundary or background layer \\
    Atmospheric & Subtle lighting for mood creation \\
    Neutral & Balanced fill without drawing attention \\
    Attention & Guides audience attention \\
    Dominant & Strongest visual focus \\
    \bottomrule
  \end{tabular}
\end{table}

Rhythmic tempo refers to the pace and regularity of lighting transitions, such as intensity changes, color shifts, and beam movements. It describes how quickly the lighting appears to change over time, often in response to musical rhythm or narrative pacing. The individual levels are defined in Table~\ref{tab:tempo}.
\begin{table}[t]
\centering
\caption{\textbf{Rhythmic Tempo Attribute Levels.} Each level defines a transition pace for preference specification.}
\label{tab:tempo}
\begin{tabular}{cc}
\toprule
\textbf{Level}  & \textbf{Description} \\
\midrule
Static  & Minimal change, nearly static \\
Slow & Gentle \& Sparse transitions, rare rhythmic cues \\
Moderate & Regular tempo, structured but not fast \\
Fast & Frequent changes, strong rhythmic presence \\
Rapid & Intense flashes with strong beats \\
\bottomrule
\end{tabular}
\end{table}

\subsection{PAMoE Gate Sensitivity }
Across 33 available preference labels, the mean weights of the four PAMoE experts are $[0.308,0.239,0.192,0.261]$, with standard deviations $[0.089,0.063,0.042,0.096]$. Experts 0, 3, and 1 are dominant for 16, 13, and 4 labels, respectively. Preference--expert correlations are structured: Color correlates most with Expert~3 ($0.471$), Brightness with Expert~1 ($0.631$), Zone with Expert~2 ($0.513$), and Rhythm with Expert~1 ($0.375$). In six existing preference-labeled outputs, Rhythm also correlates with motion ratio ($0.700$) and attribute entropy ($0.691$). The gate analysis directly demonstrates preference sensitivity; the output correlations are observational rather than a controlled fixed-music causal sweep because the latter assets were unavailable for this auxiliary analysis. 
Experts 0--3 are identifiers for symmetric learned branches, not predefined Color, Brightness, Zone, and Rhythm experts. The reported correlations describe emergent associations and do not establish exclusive specialization. No auxiliary specialization loss or hard expert assignment is used.

\section{Patching Mechanism Analysis}
\begin{figure*}[!htbp]
  \centering
  \includegraphics[width=\linewidth]{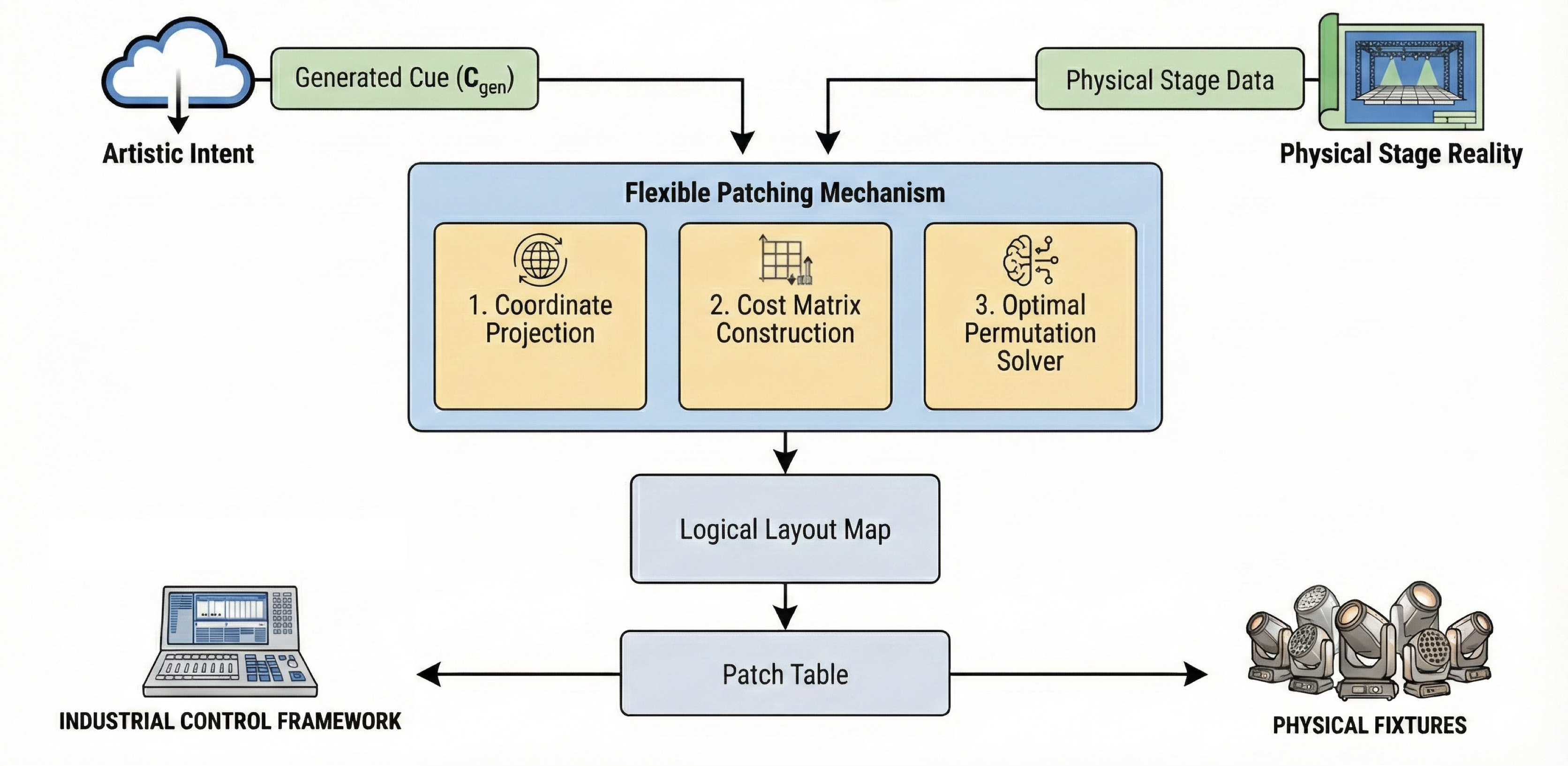}
  \caption{\textbf{Patching Mechanism Overview.} Retrieved artistic cues are retargeted to compatible fixtures through a logical layout map and an industrial-console patch table used for dispatch to the console.}
  \Description{A schematic showing retrieved lighting cues being retargeted through compatibility checks, a logical layout map, and a patch table before console control.}
  \label{fig:patching}
\end{figure*}

\label{D}
\label{sec:patching}

Prior to formalizing the fixture patching procedure, it is imperative to delineate the architectural abstractions and hardware-software synchronization protocols governing stage configuration storage in industrial control frameworks (e.g., GrandMA2).
Each fixture profile contains a \textbf{DMX Map}, a \textbf{Spatial Component} $\mathbf{p}_{norm}=(x,y,z)$, and a \textbf{Functional Component}
$\mathbf{c}_{req}=[c_{pan},c_{tilt},c_{rgb},c_{cmy},c_{gobo},\dots]$
that encodes the required hardware features. Together, these components form a \textbf{Logical Layout Map} that stores stage-specific device details for retargeting on the destination stage.

Specifically, a DMX Map defines which channel controls each physical motor or LED within a fixture. The Spatial Component is a normalized coordinate vector that serves as the geometric anchor for an individual fixture in a three-dimensional Cartesian environment. This vector maps the fixture's physical installation point relative to a predefined origin.
The Functional Component serves as a mathematical abstraction that encodes the essential hardware specifications of a device into a requirement vector. Within this vector, each element represents a discrete physical attribute or mechanical capability. For instance, the variable $c_{pan}$ signifies the fixture's ability to process horizontal rotation commands; in a practical environment, this dictates whether the hardware possesses the motors and circuitry required to perform panning movements. This logic applies consistently across the entire vector, where each subsequent element corresponds to a specific hardware-enabled function.

Our flexible patching mechanism accepts an AuraLuxMuse retrieved/adapted cue $\mathbf{C}_{out}=\langle\mathbf{p}_{norm},\mathbf{c}_{req},Attribute,val\rangle$ and a digital twin parsed from the MVR (My Virtual Rig) standard. The digital--physical address mapping defines a \textbf{Patch Table} for DMX fixture control. Musilux covers the fixture categories in Fig.~\ref{fig:lights}; compatibility with a target stage is explicitly checked through $\mathbf{c}_{req}$ rather than assumed. \footnote{MVR is an open-source XML-based file format developed to exchange stage descriptions between CAD software and lighting consoles.}
For a different stage setup, retargeting is formulated as a Linear Sum Assignment problem. A conflict occurs when multiple retrieved commands assign different values to the same fixture--attribute pair in one execution window. Real-world fixture coordinates are projected into normalized space $\hat{\mathbf{x}}$, and $C_{ij}$ combines spatial Euclidean distance with hardware/attribute mismatch penalties. The Kuhn--Munkres solution $\pi^*$ assigns commands to spatially and functionally compatible slots before console dispatch. In a deliberately conflict-heavy case with 148 fixture--attribute commands, 94 collisions occur before assignment and 0 remain afterward; attribute compatibility is preserved. The low same-fixture preservation rate in this stress test reflects the intentionally dense collision pattern. Normal exports rarely require this much reassignment because retrieved sequences usually control complementary spatial components. 

\section{Ablation Studies}
\label{E}
We perform four groups of ablation studies to assess the contributions, structures, and design choices of the AuraLuxMuse components.
\paragraph{Training Cost and Failure Signatures}
On one RTX 4090, LAMP requires approximately 50--60 minutes for 500 epochs. Joint LAMP and PAMoE training takes about 1 hour, the TCN Metadata Predictor takes 10--20 minutes, and the full pipeline takes approximately 1.0--1.5 hours. Unsuitable contrastive settings slow or destabilize alignment-loss reduction, while unsuitable density-prediction settings destabilize the cue-count loss; these effects later weaken temporal alignment or visual integrity.

\subsection{Structural Variants of Modules} 
\label{E1}
\noindent \paragraph{1. Ablation Study on LAMP Backbone Structures}
We begin by investigating architectural variants of LAMP to identify the most effective configuration. As shown in Table~\ref{tab:ablation2}, the combination of a VGGish-based music encoder with MLP layers and a transformer-based lighting encoder yields the most favorable performance.
\begin{table}[t]
  \centering
  \caption{\textbf{Ablation of LAMP Architectures.} Contrastive loss is reported after 500 training epochs of model optimization.}
  \label{tab:ablation2}
  \small
  \begin{tabular}{ccc}
    \toprule
    \makecell[c]{\textbf{Music Encoder}\\\textbf{/ Cue Encoder}}&\textbf{Transformer}&\textbf{ResNet}\\
    \midrule
    \textbf{VGGish + MLP} & \textbf{0.699($\downarrow$)} & 2.015\\
    \textbf{VGGish + ResNet} & 2.181 & 3.926  \\
  \bottomrule
\end{tabular}
\end{table}

\noindent \paragraph{2. Ablation Study on PAMoE Influence}

Subsequently, we assess PAMoE's impact and the combination of soft and hard alignment through ablations with and without these components. As shown in Table~\ref{tab:ablation3}, incorporating PAMoE improves both objective metrics and subjective perception, highlighting the value of integrating expert-level artistic knowledge into the learning process. Moreover, soft alignment improves performance by encouraging shared latent lighting behavior across stages within the same scenario.
\begin{table}[t]
  \centering
  \caption{\textbf{Ablation of PAMoE and Soft Alignment.} All experiments use the same learning rate, weight decay, and number of epochs.}
  \label{tab:ablation3}
  \begin{tabular}{cccccc}
    \toprule
    \textbf{Model Variant}&\textbf{FID$(\downarrow)$}&$\mathbf{S_{exp}}$&$\mathbf{S_{non}}$&\textbf{EGSS}$(\uparrow)$\\
    \midrule
    \textbf{PAMoE Only} & 18.06 & 2.26 & 2.86 & 2.44  \\
    \textbf{Soft Only} & 10.97 & 2.83 & 3.47 & 3.02  \\
    \textbf{PAMoE + Soft}  & 4.89 & 3.19 & 3.45 & 3.27 \\
  \bottomrule
\end{tabular}
\end{table}

\noindent \paragraph{3. Metadata Predictor Effects}

To determine dynamically the number of retrieved lighting sequences (the $K$ value) based on musical rhythm, we formulate a cross-modal dense temporal-prediction task.
Frame-level audio features (extracted using VGGish, $D=128$) are segmented into overlapping temporal windows with a total length of $W=64$ frames and stride $S=4$.
The supervision target is the discrete count of lighting events within each window.
As shown in Table~\ref{tab:metadata}, the frame-wise MLP baseline fails under unified window-level evaluation because of severe overprediction. This result highlights that mapping isolated audio frames to sparse lighting events without a sufficient contextual receptive field accumulates errors. In contrast, our non-causal TCN-based Metadata Predictor uses symmetric padding to model preceding and subsequent context within each complete input window. It aligns more closely with the sparsity of professional stage lighting and provides the density estimate $\boldsymbol{\lambda}_s$ that guides the downstream retrieval module.

\begin{table}[t]
\centering
\caption{\textbf{Metadata Predictor Performance.} We compare window-level cue counting with a baseline MLP and our non-causal TCN.}
\label{tab:metadata}
\scriptsize
\begin{tabular}{@{}cccccc@{}}
\toprule
\textbf{Architecture} & \textbf{MAE} $\downarrow$ & \textbf{RMSE} $\downarrow$ & \textbf{R$^2$} $\uparrow$ & \textbf{Pearson $r$} $\uparrow$ & \textbf{Bias} $\downarrow$ \\ \midrule
MLP & 16.987 & 24.113 & -368.715 & 0.061 & +16.973 \\
TCN Density & \textbf{0.343} & \textbf{0.449} & \textbf{0.872} & \textbf{0.934} & \textbf{+0.017} \\ \bottomrule
\end{tabular}

\end{table}

%\label{sec:lamp_ablation}
\subsection{Baseline Architecture Comparison} 
\label{E2}

\noindent \paragraph{1. Contrastive Learning Baselines}

To evaluate the effectiveness of our cross-modal alignment, we benchmark AuraLuxMuse against three representative representation learning architectures adapted for music-to-lighting retrieval:
(1) CLAP~\cite{CLAP}, a dual-encoder architecture trained with
symmetric InfoNCE contrastive loss to align audio
and text representations in a shared embedding space;
(2) SLAP~\cite{SLAP}, which adopts a Siamese architecture that
learns audio–language correspondence without explicit
negative samples via self-supervised representation matching;
(3) AudioCLIP~\cite{audioclip}, which extends CLIP-style contrastive learning
to three modalities (audio, image, and text) and learns
a unified embedding space through multi-way cross-modal
contrastive objectives. As shown in Table~\ref{tab:contrastive}, the adapted generic contrastive models struggle with the discrete, physically constrained stage-lighting space. AudioCLIP is the strongest of these baselines, while AuraLuxMuse obtains the best semantic-retrieval and expert-guided subjective scores under this evaluation. 

\begin{table}[t]
\centering
\caption{\textbf{Comparison with Contrastive Baselines.} AuraLuxMuse obtains the highest Precision@5 and EGSS under the reported setup.}
\label{tab:contrastive}
\begin{tabular}{@{}ccccc@{}}
\toprule
\textbf{Architecture} & \textbf{FID} $\downarrow$ & \textbf{EGSS} $\uparrow$ & \textbf{Precision@5} $\uparrow$ & \textbf{mAP} $\uparrow$ \\ \midrule
CLAP & 5.06 & 2.52 & 0.11 & 0.043 \\
SLAP & 7.20 & 2.11 & 0.02 & 0.085 \\
AudioCLIP & 5.21 & 2.18 & 0.13 & 0.05 \\ \midrule
\textbf{AuraLuxMuse} & 4.89 & \textbf{3.27} & \textbf{0.170} & \textbf{0.100} \\ \bottomrule
\end{tabular}
\end{table}

\noindent \paragraph{2. End-to-End Generative Baselines}

Because lighting-cue generation has not previously been formulated as a sequence-to-sequence task, we introduce a custom CueTokenizer to bridge continuous VGGish audio features and structured lighting controls. It converts lighting events into token sequences by quantizing continuous parameters into 256 bins and mapping categorical attributes to an NLP-like vocabulary. Using this formulation, we train four autoregressive baselines with token-level cross-entropy and teacher forcing: lightweight adaptations of BART~\cite{bart} and T5~\cite{T5}, which inject linearly projected audio frames as encoder embeddings; a domain-specific SpecTNT~\cite{transformermusic} surrogate with a lightweight causal Transformer decoder; and an Autoformer~\cite{autoformer} surrogate with a higher-capacity sequence-modeling variant.
As shown in Table~\ref{tab:generative}, the tested autoregressive baselines have much higher inference latency and can emit invalid attribute combinations. AuraLuxMuse instead retrieves expert-authored sequences and patches them to compatible fixtures. This design preserves corpus-level cue validity and yields 7.65-second average inference under the reported setup; final physical validity still depends on the compatibility checks and conflict-resolution procedure. 
\begin{table}[t]
\centering
\caption{\textbf{Comparison with Generative Baselines.} Against four sequence-to-sequence models, AuraLuxMuse achieves lower inference latency and a higher EGSS under the reported setup.}
\label{tab:generative}
\begin{tabular}{@{}cccc@{}}
\toprule
\textbf{Architecture} & \textbf{FID} & \textbf{EGSS} $\uparrow$ & \textbf{Inference Time (s/file)} $\downarrow$ \\ \midrule
T5                 & 4.28        & 1.79                                     & 25.74                                         \\
BART              & 4.21        & 1.89                                     & 35.20                                         \\
Autoformer        & 3.98       & 1.24                                     & 207.27                                        \\
SpecTNT          & 3.95        & 1.24                                     & 150.18                                        \\ \midrule
\textbf{AuraLuxMuse} & 4.89                       & \textbf{3.27}   & \textbf{7.65}                                 \\ \bottomrule
\end{tabular}
\end{table}

The generative baselines perform poorly because attributes and values are highly correlated, while discrete tokenization produces a combinatorial number of possible token sequences. Without explicit retrieval guidance, the autoregressive models struggle to learn the dependencies and constraints of stage-lighting design, resulting in incoherent outputs and lower subjective scores.

\subsection{Retrieval Baseline Diagnostics}
The rule-based baseline partitions the cue corpus with coarse audio-duration and complexity heuristics and then samples compatible professional cues. It is physically executable and interpretable but does not learn music--lighting correspondence. Random retrieval shares the Metadata Predictor's event timing and samples from the same expert-authored, console-valid corpus; this explains why it is a strong corpus-prior lower bound rather than a zero-quality baseline. Its weakness is semantic alignment: random retrieval scores 2.70 in narrativity versus 3.64 for similarity retrieval, while visual consistency is closer (3.30 versus 3.55). Auxiliary cue statistics support this distinction: random outputs have a substantially larger mean cue-statistic distance from similarity retrieval than the rule-based outputs reported and analyzed above. 

\subsection{Delicate Component Mechanisms}
\label{E3}

\noindent \paragraph{1. Ablation Study on $K$-Value Design}
The symbol $K$ denotes the number of retrieved sequences computed by integrating the predicted event density. All top-$K$ sequences are considered because an individual sequence may control only one stage region; conflicts are resolved before dispatch. Table~\ref{tab:k} compares adaptive $K$ with fixed values, and the adaptive setting obtains the best reported objective and subjective scores among those tested configurations. 
\begin{table}[t]
\centering
\caption{\textbf{Ablation Study: Static vs. Dynamic $K$ Values.} Dynamic $K$ yields a lower FID and a higher EGSS than either static setting.}

\begin{tabular}{cccc}

\toprule
\textbf{Strategy} & \textbf{$K$} & \textbf{FID} ($\downarrow$) & \textbf{EGSS} ($\uparrow$) \\
\midrule
Fixed Small & 1 & 8.42 & 2.85 \\
Fixed Large & 10 & 9.82 & 3.02 \\
\textbf{Adaptive} & \textbf{Dynamic} & \textbf{4.89} & \textbf{3.27} \\
\bottomrule
\end{tabular}

\label{tab:k}
\end{table}

\noindent \paragraph{2. Ablation Study on Soft Alignment and Weights}
To investigate the contribution of the soft alignment mechanism in our contrastive learning objective, we perform an ablation study on the soft positive weight, denoted as $w$. While fixing the standard hard alignment weight at 1.0, we systematically sweep $w \in \{0.1, 0.3, 0.5, 0.7, 0.9\}$.
The introduction of soft alignment is highly beneficial for the stage lighting domain; unlike standard text-image pairs, multiple visually distinct lighting cues can be aesthetically compatible with the same musical rhythm. By assigning a partial target probability ($w$) to these style-consistent samples, the soft alignment loss provides a smoothed learning signal that prevents the model from overly penalizing valid stylistic variations, thereby enriching the semantic continuity of the shared embedding space.
As shown in Table~\ref{tab:weights}, setting $w=0.3$ provides the best configuration. When $w$ is too small (e.g., 0.1), strict hard alignment dominates and the model does not fully exploit latent stylistic correlations. Conversely, increasing the soft weight to $w \geq 0.5$ causes soft positives to overpower the exact ground-truth anchors. This excessive smoothing blurs discriminative boundaries in the feature space, as evidenced by a substantial drop in retrieval recall and an increase in FID to 61.21 at $w=0.7$. At $w=0.3$, the network achieves the highest overall retrieval accuracy while preserving a stable representation manifold.

\begin{table*}[ht]
\centering
\caption{\textbf{Ablation on Soft Alignment Weight ($w$).} We evaluate the impact of the soft-alignment weight $w$ on cross-modal retrieval performance and Fr\'echet Inception Distance (FID). The hard-alignment weight is fixed at 1.0. Results indicate that $w=0.3$ yields the best balance, achieving the strongest overall retrieval metrics by leveraging stylistic similarities without blurring discriminative boundaries.}
\label{tab:weights}
\begin{tabular}{@{}lcccccc@{}}
\toprule
\textbf{Soft Weight ($w$)} & \textbf{Avg FID}  $\downarrow$ & \textbf{R@1} $\uparrow$ & \textbf{R@5} $\uparrow$ & \textbf{R@10} $\uparrow$ & \textbf{MRR} $\uparrow$ & \textbf{nDCG} $\uparrow$ \\ \midrule
$w = 0.1$ & 11.85 & 0.045 & 0.223 & 0.338 & 0.144 & 0.303 \\
$\mathbf{w = 0.3}$ &  \textbf{4.89} & 0.046 & \textbf{0.249} & \textbf{0.374} & \textbf{0.149} & \textbf{0.309} \\
$w = 0.5$ &10.10 & \textbf{0.062} & 0.202 & 0.280 & 0.143 & 0.300 \\
$w = 0.7$ & 61.21 & 0.033 & 0.159 & 0.301 & 0.118 & 0.279 \\
$w = 0.9$ & 47.09 & 0.020 & 0.127 & 0.258 & 0.097 & 0.261 \\ \bottomrule
\end{tabular}
\end{table*}

\noindent \paragraph{3. Ablation Study on Lighting Embedding Strategies}
We evaluate two cue embedding strategies: attribute mapping, which assigns predefined indices to categorical lighting attributes, and one-hot encoding, which uses high-dimensional binary vectors. As shown in Table~\ref{tab:ablation1}, attribute mapping consistently outperforms one-hot encoding across objective metrics and subjective evaluations.
\begin{table}[t]
  \centering
  \caption{\textbf{Ablation of Cue-Embedding Strategies.} The architecture is fixed, and only the embedding strategy varies.}
  \label{tab:ablation1}
  \begin{tabular}{ccccc}
    \toprule
    \makecell[c]{\textbf{Embedding}}&\textbf{FID$(\downarrow)$}&$\mathbf{S_{exp}}$&$\mathbf{S_{non}}$&\textbf{EGSS}$(\uparrow)$\\
    \midrule
    \textbf{One Hot}  & 12.24 & 2.51 & 3.29 & 2.74\\
    \textbf{Attribute Map}  & 4.89 & 3.19 & 3.45 & 3.27\\
  \bottomrule
\end{tabular}
\end{table}

%\label{sec:pamoe_ablation}
\subsection{Metric Validity and Reliability} 
\label{E4}
\begin{figure}[t]
  \centering
  \includegraphics[width=\columnwidth]{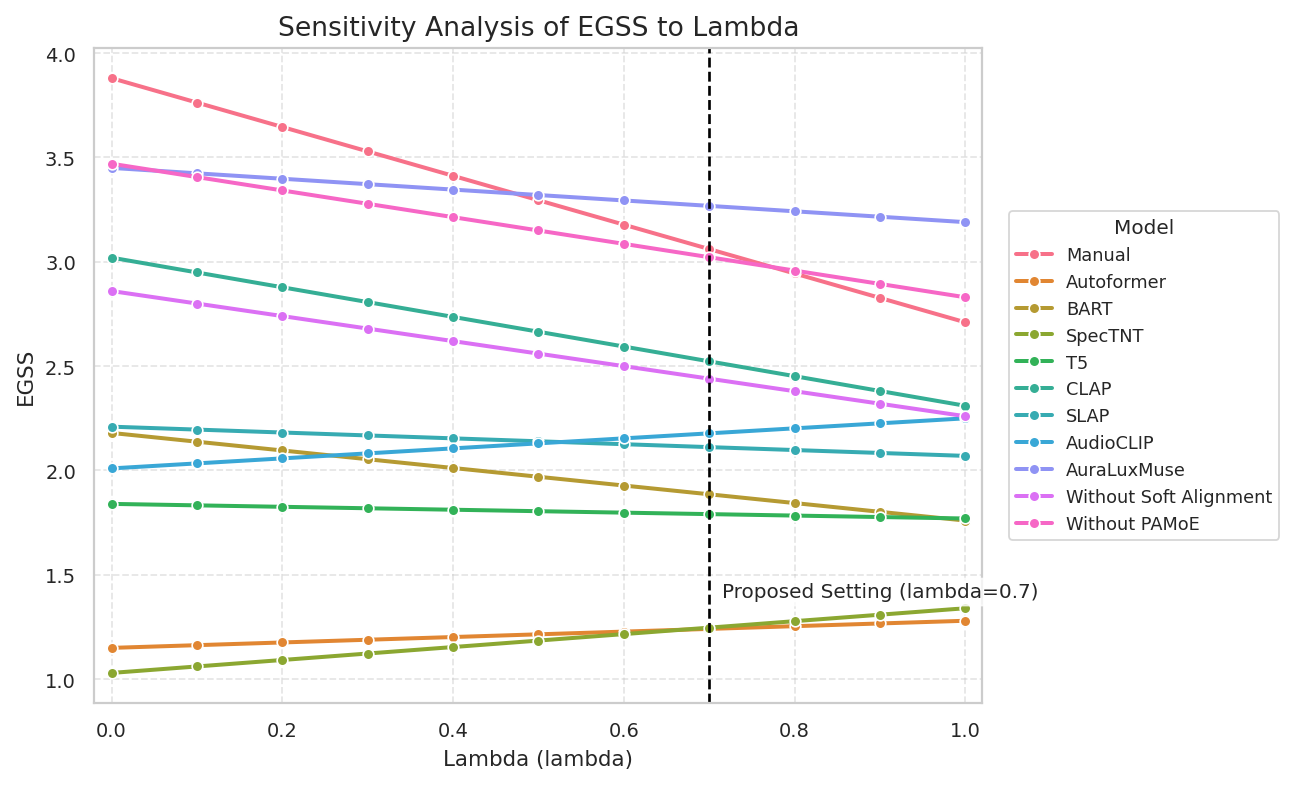}
  \caption{\textbf{Sensitivity analysis of the EGSS metric with respect to $\lambda$.} The parameter $\lambda$ controls the weight assigned to expert preferences in the subjective evaluation process. The manual lighting design serves as an approximate upper bound. AuraLuxMuse consistently outperforms all automated baselines across the full $\lambda$ range, demonstrating strong robustness to this hyperparameter. The vertical dashed line indicates our selected default configuration ($\lambda=0.7$).}
  \Description{A line chart showing EGSS for manual design and automated baselines as the expert-preference weight lambda varies, with a dashed line marking lambda equal to 0.7.}
  \label{fig:sensitivity}
\end{figure}

\noindent \paragraph{1. Sensitivity Analysis}
To validate our evaluation protocol, we investigate the sensitivity of the Expert-Guided Subjective Score (EGSS) to the expert-weight parameter $\lambda$. As illustrated in Fig.~\ref{fig:sensitivity}, varying $\lambda$ systematically adjusts the emphasis placed on professional lighting principles relative to general visual perception. The manual design serves as an approximate upper bound rather than a competing baseline. Across the entire $\lambda$ range, AuraLuxMuse outperforms the tested automated baselines. We adopt $\lambda=0.7$ as the default configuration because it balances expert alignment with general aesthetic perception under the reported study design.

\noindent \paragraph{2. Correlation Analysis between FID and EGSS}
We examine FID and EGSS across ten automated model variants. Pearson correlation is positive but not significant ($r=0.3955$, $p=0.2579$), whereas Spearman correlation is positive and significant in this sample ($\rho=0.6848$, $p=0.0289$). Because this analysis has only ten variants, excludes manual design, and evaluates FID in one learned feature space, it indicates exploratory rank association rather than robust metric validity. FID and EGSS are therefore complementary diagnostics, not interchangeable aesthetic measures.

\label{sec:embedding_ablation}

\section{Stage Lighting Environments}
\label{F}

\begin{figure*}[ht]
  \centering
  \includegraphics[width=0.9\linewidth]{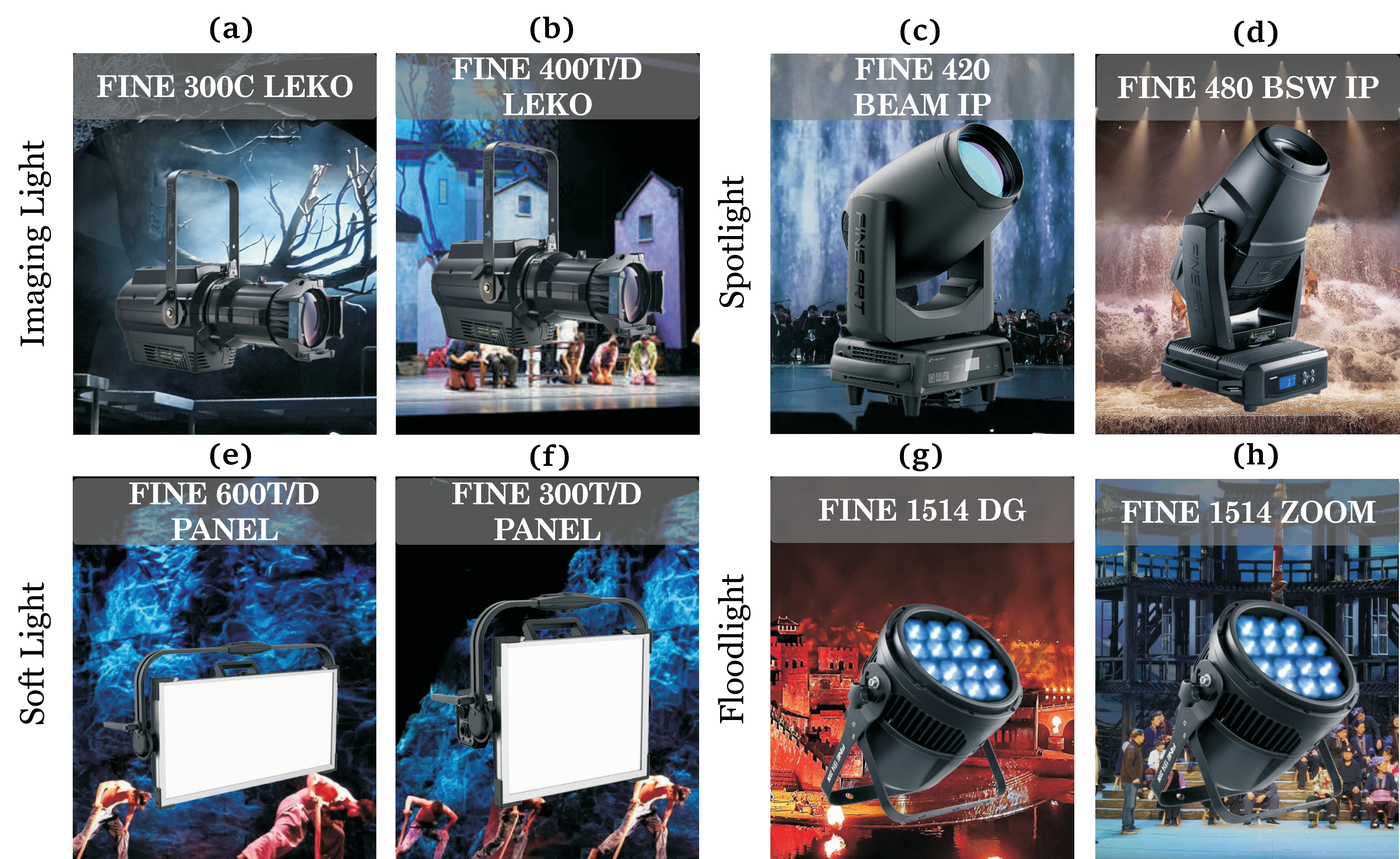}
  \caption{\textbf{Examples of Stage Lighting Fixtures:} (a-b) Imaging Lights: FINE 300C LEKO and FINE 400T/D LEKO for sharp projection and patterned illumination; (c-d) Spotlights: FINE 420 BEAM IP and FINE 480 BSW IP for dynamic focus and concentrated beams; (e-f) Soft Lights: FINE 600T/D PANEL and FINE 300T/D PANEL for diffused illumination and even coverage; (g-h) Floodlights: FINE 1514 DG and FINE 1514 ZOOM for broad lighting and wide effects.}
  \Description{An eight-panel image sheet showing imaging lights, spotlights, soft lights, and floodlights used in the stage-lighting setup.}
  \label{fig:lights}
\end{figure*}

\subsection{Software}
\label{F1}

For high-dimensional parameter modulation, sequence orchestration, and real-time output synchronization, we employ GrandMA2 as our primary lighting-control interface. For synthetic data generation and previsualization, MA3D renders complex volumetric lighting effects in a simulated environment. The system also supports hardware-in-the-loop execution: control signals generated through GrandMA2 are mapped identically to both the virtual engine and physical stage luminaires. This unified protocol ensures spatiotemporal consistency between simulated and real-world lighting distributions, providing a framework for cross-domain validation.

GrandMA2 offers a robust, scalable solution for professional stage lighting control, integrating high-performance hardware with versatile software to support a wide range of production needs. The system supports seamless show file management, real-time cue execution, and parameter expansion across a unified control interface. The GrandMA2 software enables rich programming functionality including presets, effects, and time-synchronized lighting behaviors, with consistent UI across hardware configurations. Designed for demanding live environments, GrandMA2 provides reliable performance for productions ranging from small installations to arena-scale events.

MA3D is a high-fidelity real-time rendering and visualization engine designed for complex lighting configurations. It serves as a digital twin of the physical stage, enabling simulation of volumetric light beams, non-Lambertian shadows, and dynamic spectral distributions across three-dimensional occluders. The engine maintains bidirectional data synchronization with GrandMA2 through a high-speed network interface, ensuring that virtual lighting states are spatiotemporally aligned with hardware-level DMX output.

\subsection{Lighting Fixtures}
\label{F2}

\begin{figure}[ht]
    \centering
    \includegraphics[width=\linewidth]{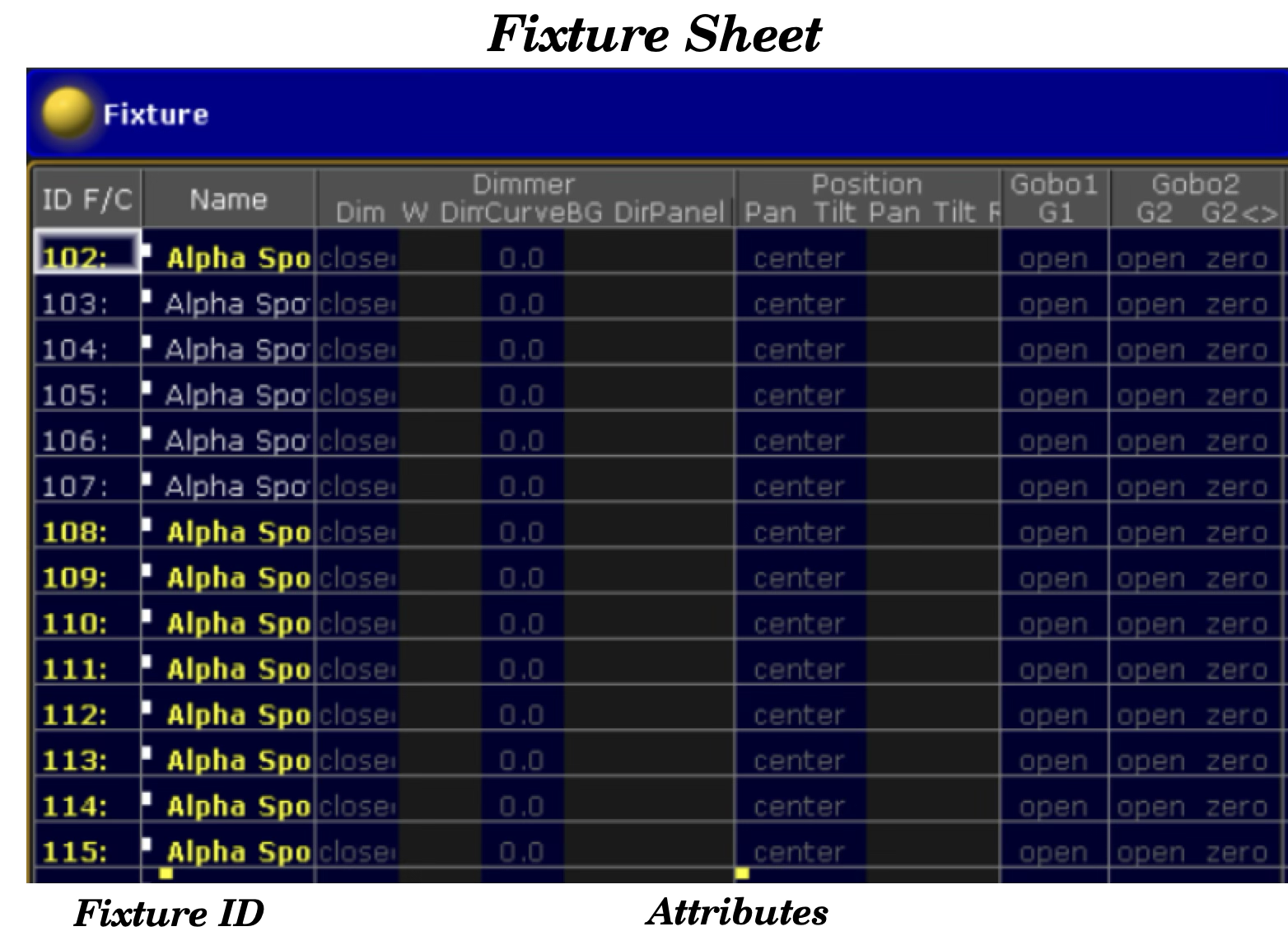}
    \caption{\textbf{Fixture Parameter Visualization.} The Fixture Sheet records instantaneous lighting-infrastructure states. Its fields match the functional components in Section~\ref{sec:patching}, enabling precise state tracking during acquisition.}
    \Description{A fixture-sheet table listing the instantaneous state parameters of the lighting infrastructure for data acquisition and tracking.}
    \label{fig:fixture}
\end{figure}

\begin{figure*}[ht]
    \centering
    \includegraphics[width=\linewidth]{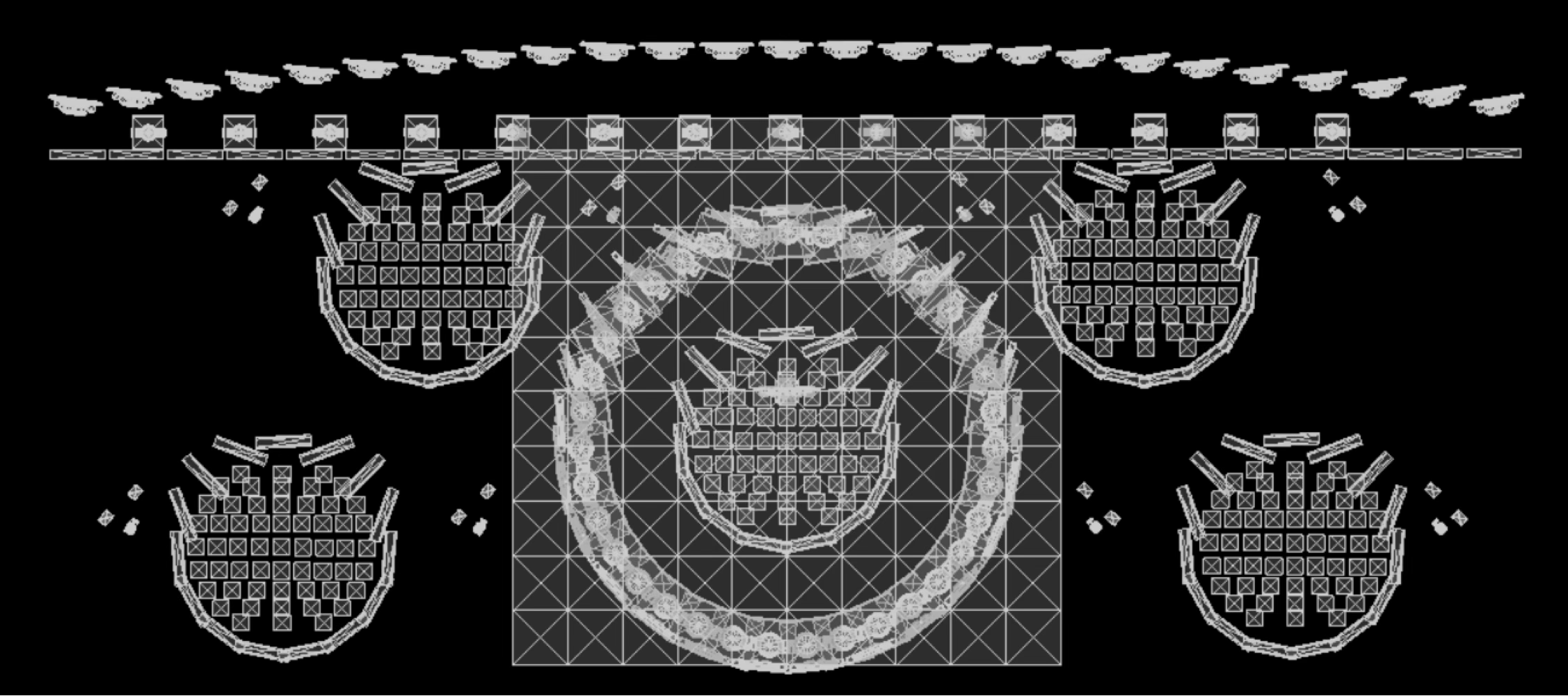}
    \caption{\textbf{Software-Defined Spatial Manifold: Top-down View.} The top-down orthographic projection of the experimental environment as registered within the GrandMA2 configuration space. Each luminaire instance and structural element is precisely localized within a unified global coordinate system, providing the necessary extrinsic parameters for spatiotemporal alignment. This geometric blueprint serves as the architectural ground truth, ensuring that synthesized radiance fields are spatially consistent with the physical acquisition environment.}
    \Description{A top-down orthographic diagram of the experimental stage showing the registered positions of luminaires and structural elements in a shared coordinate system.}
    \label{fig:top}
\end{figure*}

\begin{figure*}[ht]
    \centering
    \includegraphics[width=\linewidth]{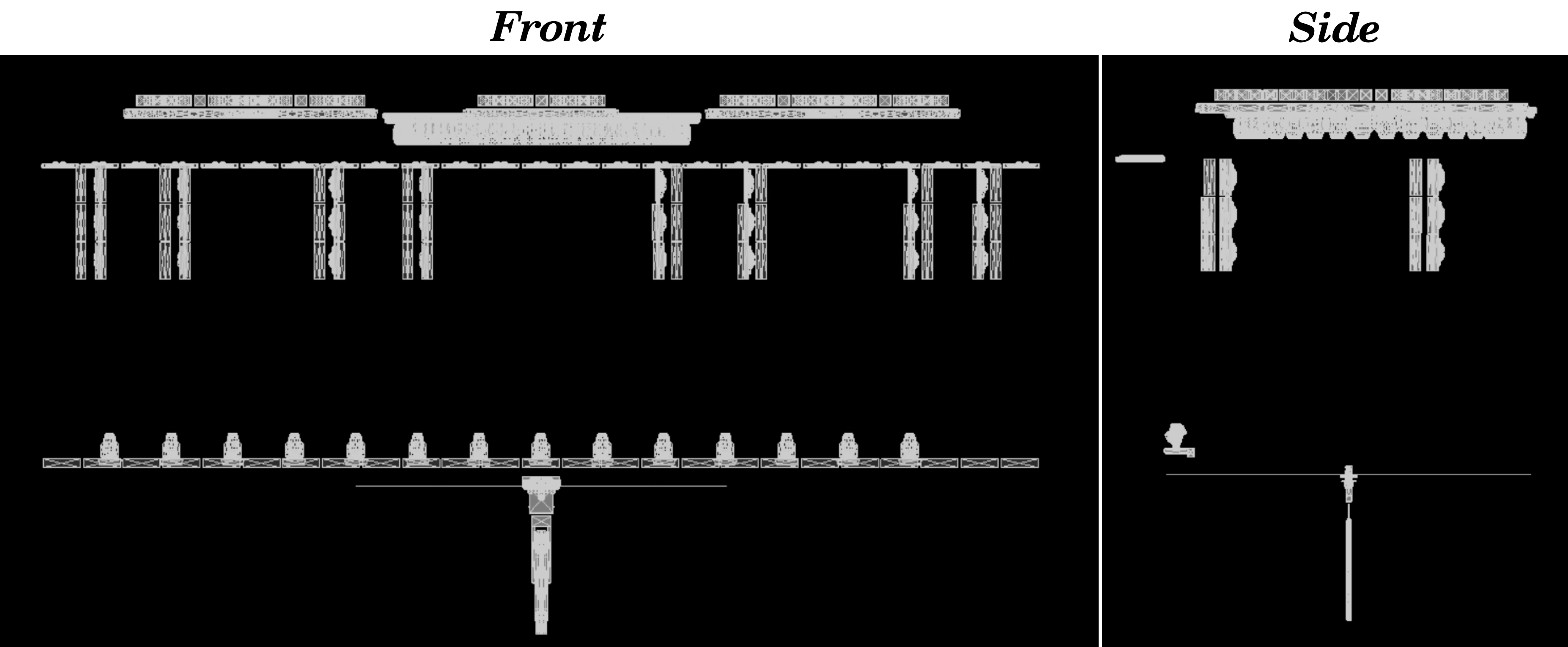}
    \caption{\textbf{Software-Defined Spatial Manifold: Front and Side Views.} Registered orthographic projections map fixture elevations and lateral offsets, providing ground truth for spatial registration.}
    \Description{Front and side orthographic diagrams of the experimental stage showing fixture elevations and lateral offsets used for spatial registration.}
    \label{fig:front_side}
\end{figure*}

\begin{figure*}[!htbp]
    \centering
    \includegraphics[width=\linewidth]{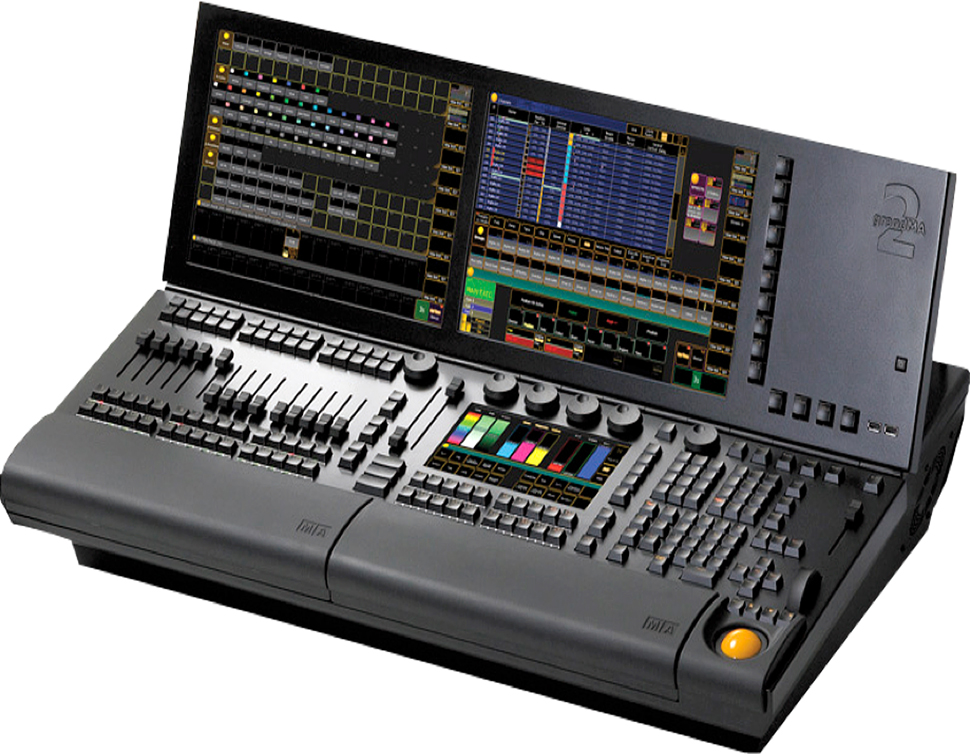}
    \caption{\textbf{Example of a GrandMA2 Console.} The professional console provides robust stage-lighting control.}
    \Description{A photograph of a GrandMA2 professional lighting console with multiple faders, displays, and control buttons.}
    \label{fig:console}
\end{figure*}

Stage-lighting fixtures can be classified by the morphology of their emitted light into four classes: imaging lights, spotlights, soft lights, and floodlights. Fig.~\ref{fig:lights} shows the fixture models represented in Musilux. Each fixture type serves a distinct purpose in shaping the visual atmosphere of a performance and offers different degrees of beam control, diffusion, and intensity.

Imaging lights, often equipped with advanced optical systems, allow precise projection of patterns, gobos, or textured effects. Spotlights use focused, directional beams to highlight performers or key stage elements with sharp, well-defined edges. In contrast, soft lights produce diffuse, even illumination that minimizes harsh shadows, making them suitable for naturalistic or ambient conditions. Floodlights deliver broad, uniform coverage and are frequently used to wash large stage areas or backdrops. These four fixture classes cover the professional-stage configurations represented in Musilux.

Fig.~\ref{fig:fixture} depicts fixture parameters stored in the stage-lighting control interface. The interface maps hardware IDs to lighting and spatial-transform vectors, recording extrinsic parameters alongside the multimodal data. This deterministic mapping from logical control to physical fixture behavior supports tasks such as inverse rendering and scene decomposition.

\begin{figure*}[ht]
    \centering
    \includegraphics[width=\linewidth]{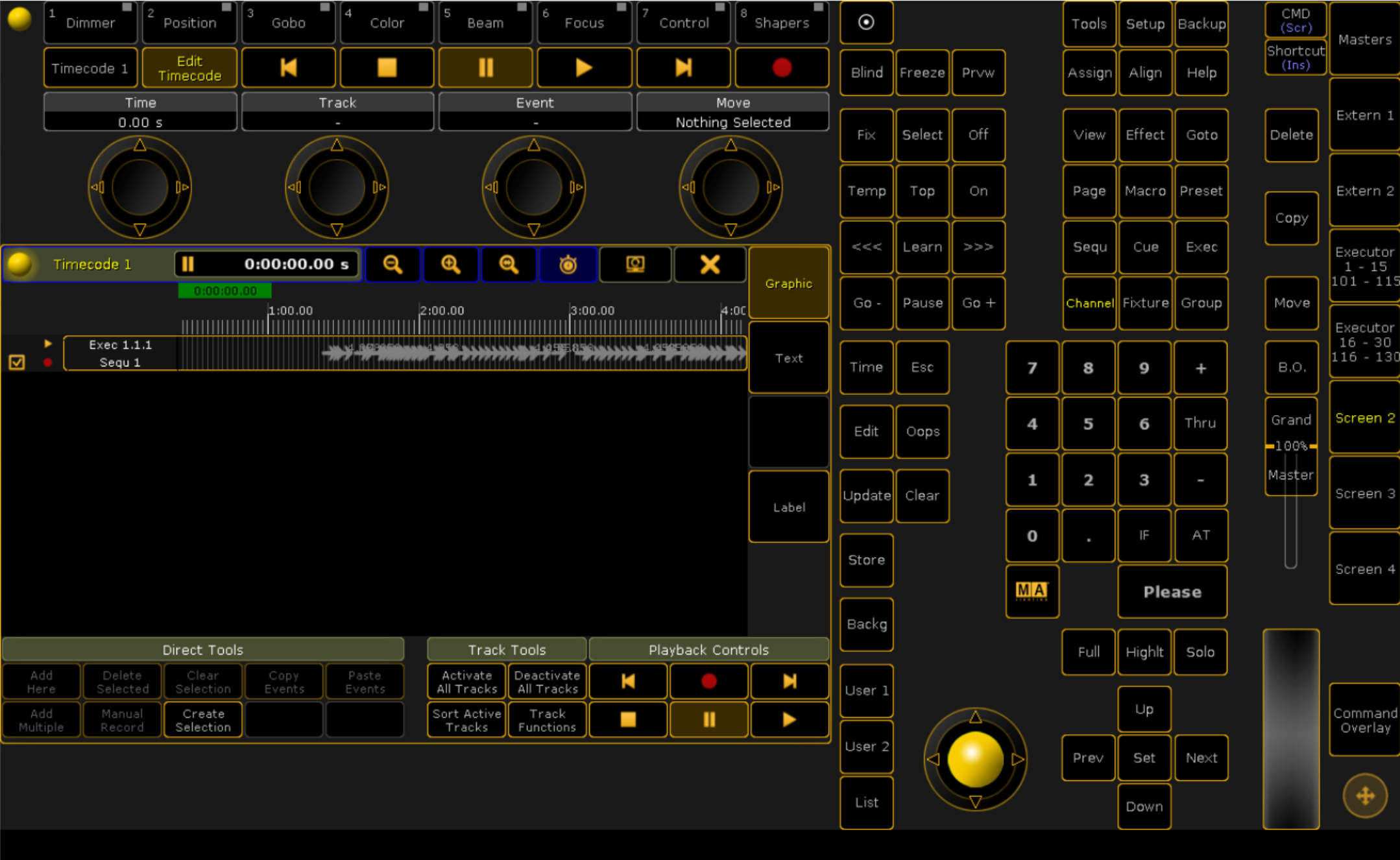}
    \caption{\textbf{GrandMA2 Software Interface.} The software provides real-time visualization, cue management, and timeline editing.}
    \Description{A screenshot of the GrandMA2 software interface showing a stage preview, cue controls, and timeline editing panels.}
    \label{fig:interface}
\end{figure*}

\subsection{Stage Configurations}
\label{F3}

In addition to software and fixtures, the stage provides the physical environment in which lighting operates. A stage configuration is stored from multiple perspectives, including two-dimensional geometric information established through three orthographic views in the industrial design process.
To establish a robust mapping between the virtual simulation and the physical capture space, the spatial configuration within the GrandMA2 ecosystem contains complex geometric occluders. As depicted in Fig.~\ref{fig:top} (Top View) and Fig.~\ref{fig:front_side} (Front \& Side View), this integrated representation facilitates the precise calibration of light transport paths and ensures that the extrinsic properties of the illumination sources are preserved across both synthetic and real-world domains.

The stage environment serves as a persistent spatial repository in which the extrinsic parameters of heterogeneous fixtures are represented in a unified three-dimensional coordinate system. These geometric priors allow the architectural configuration to be serialized in GrandMA2 show files and persistent data archives. This hierarchical representation supports rapid initialization of complex lighting configurations and import of existing stage-lighting programs. The serialized format also preserves the spatial distribution of the lighting infrastructure across project instances, supporting consistent multimodal data acquisition.

\subsection{Stage Lighting Console}
\label{F4}

Fig.~\ref{fig:console} illustrates the physical control surface used to orchestrate the lighting infrastructure. GrandMA2 provides a structured framework for temporal sequencing, per-fixture mapping, and low-level parameter modulation. Coupling the hardware surface with the software environment enables control of complex lighting trajectories. This configuration supports hardware-in-the-loop actuation, allowing designers to manipulate real-world stage lighting and refine its state in real time through a tactile control surface.

The full-size console supports 8,192 parameters natively and scales up to 65,536 via MA Network Processing Units (NPUs), while the light and ultra-light versions offer 4,096 parameters, expandable similarly. Each console features multi-touch displays, extensive DMX I/O (up to 6 outputs), MIDI/SMPTE integration, and ethernet-based protocols such as Art-Net and sACN.

\begin{figure*}[htbp]
    \centering
    \includegraphics[width=\linewidth]{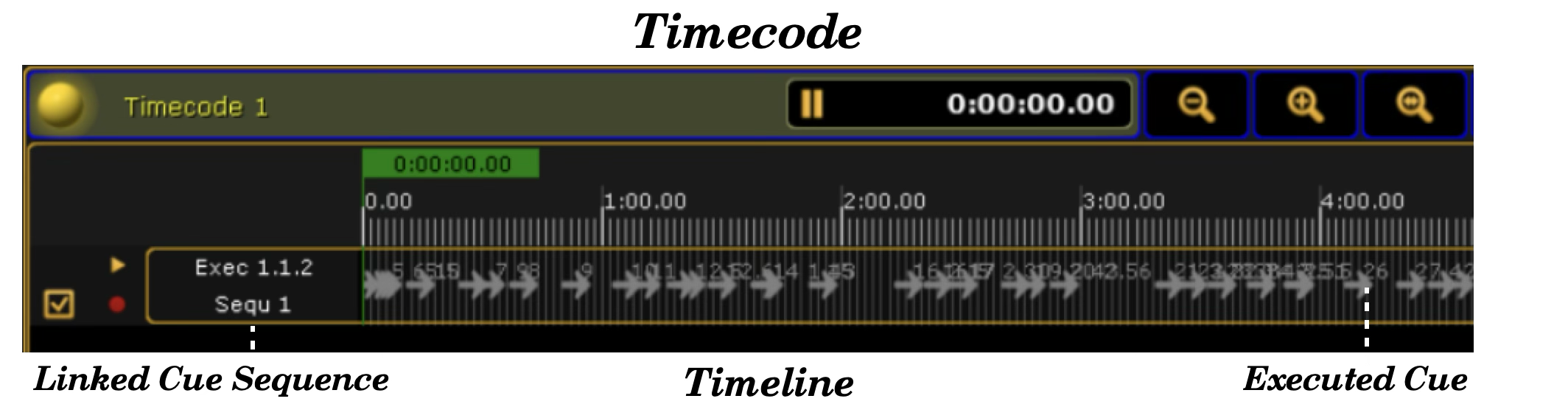}
    \caption{\textbf{Timecode Encoding in GrandMA2.} The interface precisely synchronizes multimodal lighting triggers; each marker denotes a discrete cue execution.}
    \Description{A screenshot of a GrandMA2 timecode timeline with markers indicating discrete lighting-cue executions.}
    \label{fig:timecode}
\end{figure*}

\begin{figure*}[ht]
    \centering
    \includegraphics[width=0.9\linewidth]{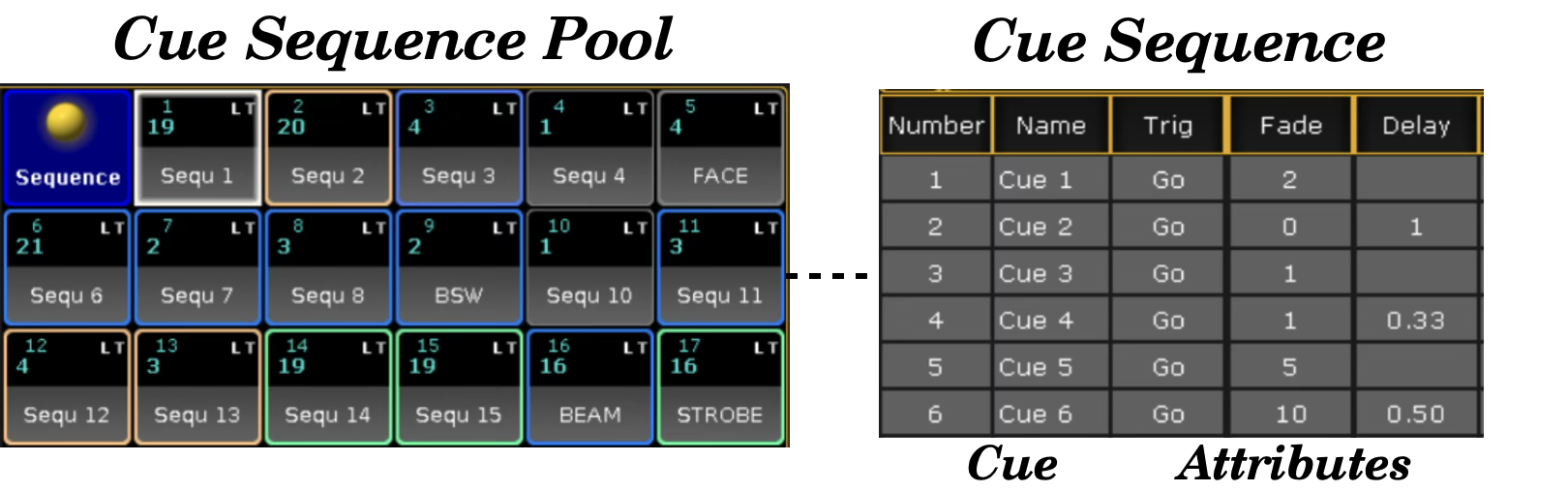}
    \caption{\textbf{Structured Parameter Space for Cue Sequences.} (Left) The Sequence Pool stores illumination trajectories. (Right) An individual Cue Sequence specifies fixture behavior through an attribute vector that includes transition duration (Fade) and temporal offset (Delay), enabling deterministic execution.}
    \Description{A two-part screenshot showing a cue-sequence repository on the left and detailed fixture-behavior parameters for one selected cue sequence on the right.}
    \label{fig:cue_sequence}
\end{figure*}

\subsection{Stage Lighting Interface}
\label{F5}

The interactive software interface is demonstrated in Fig.~\ref{fig:interface}. It is a hierarchical state-space for managing dynamic radiance environments and provides the following functional components:

\begin{itemize}
    \item \textbf{Temporal Synchronization (Timecode):} The global timecode interface in Fig.~\ref{fig:timecode} maps discrete event triggers to a high-resolution SMPTE reference.
    \item \textbf{Hierarchical Cue Sequence Representation:} The repository in Fig.~\ref{fig:cue_sequence} stores illumination events as structured Cue Sequences; each cue is a state vector that parameterizes lighting behavior.
    \item \textbf{Attribute Control Surface:} Eight encoders (Dimmer, Position, Gobo, Color, Beam, Focus, Control, and Shapers) control fixture parameters during show execution.
    \item \textbf{Tactile Command Emulation:} A 1:1 digital twin includes the Numeric Keypad, Command Buttons (e.g., Store, Update, Clear), and Execution Faders.
    \item \textbf{Navigation and Workspace Management:} Vertical tabs support rapid switching among fixture sheets and layouts.
\end{itemize}

\section{Stage Lighting Implementation Procedure}
\label{G}

This section details how to export learnable data structures from existing lighting shows and import AuraLuxMuse-generated stage lighting into real-world simulations and applications.

\subsection{Stage Lighting Input}
\label{G1}

This section describes how to export XML files from existing stage-lighting designs in GrandMA2 and convert them into readable documents for constructing Musilux.

\noindent \textbf{Step 1: Import Lighting Show.}
To begin, the show file corresponding to the specific stage lighting setup must be imported into the console. This is done by selecting the \texttt{Load Show} option, which allows the user to load the desired lighting configuration file into the GrandMA2 system.

\noindent \textbf{Step 2: Export XML Files.} 
For one sequence or timecode, execute:
\begin{itemize}
    \item \texttt{SelectDrive 4}
    \item \texttt{Export Sequence (Number of Sequence) ("File Name")}
    \item \texttt{Export Timecode (Number of Timecode) ("File Name")}
\end{itemize}

For multiple sequences or timecodes, execute:
\begin{itemize}
    \item \texttt{SelectDrive 4}
    \item \texttt{Export Sequence (Start) Thru (End) ("File Name")}
    \item \texttt{Export Timecode (Start) Thru (End) ("File Name")}
\end{itemize}

\noindent \textbf{Step 3: Data Conversion.}
The model cannot directly use the original XML stage-lighting data. Run \texttt{read\_sequence\_cue.py} and \texttt{read\_timecode.py} to extract structured information and produce JSON files used for model training.

\subsection{Stage Lighting Output}
\label{G2}
This section describes how to integrate AuraLuxMuse-generated XML files into the \textbf{GrandMA2} console for stage-lighting control during deployment on target stages.

\noindent \textbf{Step 1: File Preparation.} AuraLuxMuse generates a cue sequence XML file and a timecode XML file, which should be placed in the USB drive's \\ \texttt{\textbackslash gma2\textbackslash importexport} directory.

\noindent \textbf{Step 2: Console Setup.} Open a blank show file and configure the lighting fixtures.

\noindent \textbf{Step 3: Import Files.} Execute these macros:
\begin{itemize}
    \item \texttt{SelectDrive 4}
    \item \texttt{Import ("File Name") At Sequence 1}
    \item \texttt{Import ("File Name") At Timecode 1}
\end{itemize}
Replace "File Name" with the actual XML file names. For simplicity in subsequent explanations, the numbers are assumed to be 1, though they can be adjusted as needed.

\noindent \textbf{Step 4: Assign Sequence.} Run \texttt{Assign Sequence 1 At Executor 1} to link the cue sequence to Executor 1, which is also adjustable.

\noindent \textbf{Step 5: Bind Timecode.} Edit Timecode 1 and select the corresponding Executor 1.

\noindent \textbf{Step 6: Playback.} Click the timecode play icon to start synchronized lighting control.

This workflow enables efficient \mbox{AuraLuxMuse} deployment.

\section{Dataset Split and Generalization Analysis }
\label{H}
Training excludes the validation and test assets and contains approximately 150 minutes of paired music and lighting cues. Validation covers six styles with two held-out examples per scenario and supplies manual references for the main-paper comparison. Test uses six unseen tracks across the six styles on a stage excluded from training. This is a controlled, style-balanced check of unseen music and unseen stage layout, not a population-scale benchmark; full-length expert/non-expert scoring limits its practical sample size. A cue-statistic t-SNE analysis over 14 stage-level output vectors yields a mean pairwise z-scored distance of $5.409\pm1.746$, indicating multiple statistic regimes rather than collapse to a single template. The real-world demo stage is unseen; its nearest cue-statistic neighbors are the dance and rock stage cases. Because the t-SNE uses available cue-output vectors rather than the complete raw Musilux corpus, we treat it as supplementary descriptive evidence, not as a standalone generalization metric. 
The approximately 150-minute training portion is about 80.1\% of total Musilux duration. Validation comprises six held-out programs, two per scenario, selected for six-style coverage and the availability of human-authored references. Test comprises six additional tracks, one per style, and an unseen stage layout. These count- and duration-level descriptions are both reported because program lengths vary.

\end{document}